\RequirePackage{ifpdf}
\ifpdf 
\documentclass[pdftex]{sigma}
\else
\documentclass{sigma}
\fi

\numberwithin{equation}{section}
\numberwithin{theorem}{section}
\numberwithin{proposition}{section}
\numberwithin{lemma}{section}
\numberwithin{corollary}{section}
\numberwithin{conjecture}{section}
\numberwithin{definition}{section}
\numberwithin{example}{section}
\numberwithin{remark}{section}
\numberwithin{note}{section}

 \def\barr{\left(\begin{array}}
 \def\earr{\end{array}\right)}

 \newcommand{\R}{ {\mathbb R} }
 \newcommand{\eps}{ \varepsilon }
 \newcommand{\p}{\partial}

\begin{document}
\allowdisplaybreaks

\renewcommand{\PaperNumber}{070}

\FirstPageHeading

\ShortArticleName{On Brane Solutions Related to Non-Singular Kac--Moody Algebras}

\ArticleName{On Brane Solutions Related\\ to Non-Singular Kac--Moody Algebras}

\Author{Vladimir D. IVASHCHUK and Vitaly N. MELNIKOV}

\AuthorNameForHeading{V.D. Ivashchuk and V.N. Melnikov}

\Address{Center for Gravitation and Fundamental Metrology,
 VNIIMS,\\ 46 Ozyornaya Str., Moscow 119361, Russia \\[1mm]
  Institute of Gravitation and Cosmology,
 Peoples' Friendship University of Russia,\\
 6 Miklukho-Maklaya Str., Moscow 117198, Russia}

\Email{\href{mailto:ivashchuk@mail.ru}{ivashchuk@mail.ru}, \href{mailto:melnikov@phys.msu.ru}{melnikov@phys.msu.ru}}

\ArticleDates{Received October 01, 2008, in f\/inal form June 15, 2009;  Published online July 07, 2009}

 \Abstract{A multidimensional gravitational model containing scalar
 f\/ields  and antisymmetric forms is considered. The manifold is
 chosen in
 the form $M = M_0 \times M_1 \times \cdots \times M_n$, where
 $M_i$ are Einstein spaces ($i \geq 1$). The sigma-model approach
 and exact solutions with intersecting composite branes (e.g.~solutions with harmonic functions, $S$-brane and black brane ones)
 with intersection rules related to non-singular
 Kac--Moody (KM) algebras (e.g. hyperbolic ones) are reviewed.
 Some examples  of solutions, e.g. corresponding to hyperbolic
 KM algebras:   $H_2(q,q)$, $AE_3$, $HA_2^{(1)}$, $E_{10}$ and
 Lorentzian KM algebra  $P_{10}$ are  presented.}


\Classification{17B67; 17B81; 83E15; 83E50; 83F05; 81T30}

\section{Introduction}

 Kac--Moody (KM) Lie algebras \cite{Kac0,Moody,Kac} play a rather important
 role in dif\/ferent areas of mathe\-ma\-tical physics (see
 \cite{Kac,FS,Nik,HPS} and references therein).

 We recall  that  KM Lie algebra is a Lie algebra
 generated by the  relations \cite{Kac}
\begin{gather*}
 [h_i,h_j] =0, \qquad [e_i,f_j] = \delta_{ij} h_j, \\
 [h_i,e_j] = A_{ij} e_j, \qquad   [h_i,f_j] = -A_{ij} f_j, \\
 ({\rm ad} e_i)^{1- A_{ij}}(e_j) = 0   \qquad (i \neq j), \\
({\rm ad} f_i)^{1- A_{ij}}(f_j) = 0 \qquad (i \neq j).
\end{gather*}

  Here $A = (A_{ij})$ is a generalized  Cartan matrix,
  $i,j = 1,  \ldots, r$, and   $r$ is the rank of the KM algebra.
  It means that all  $A_{ii} =   2$;
  $A_{ij}$ for $i \neq j$ are non-positive integers and
  $A_{ij} =0 $ implies $A_{ji} =0$.

   In what follows the matrix $A$ is restricted to
   be non-degenerate (i.e.\ $\det A \neq 0$) and symmetrizable
   i.e.\ $A = B {\cal D}$, where $B$ is a symmetric matrix and
   ${\cal D}$ is an invertible diagonal matrix
   (${\cal D}$ may be chosen in such way that all
   its  entries ${\cal D}_{ii}$ are positive rational numbers \cite{Kac}).
   Here we
   do not consider singular  KM algebras with $\det A = 0$,
   e.g.\ af\/f\/ine ones. Recall that af\/f\/ine KM algebras are of much interest for
    conformal f\/ield theories, superstring theories etc.~\cite{FS,GrSW}.

    In the case when $A$ is positive def\/inite (the Euclidean case) we get ordinary f\/inite
    dimensional Lie algebras \cite{Kac,FS}. For non-Euclidean signatures of $A$
    all KM algebras are inf\/inite-dimensional. Among these the Lorentzian
    KM algebras  with pseudo-Euclidean signatures  $(-,+, \ldots, +)$
    for the Cartan matrix $A$ are of current interest,  since they contain a subclass
    of the so-called hyperbolic KM algebras widely used in modern mathematical
    physics.  Hyperbolic KM algebras are by def\/inition Lorentzian Kac--Moody
    algebras with the property that removing any node from their Dynkin
    diagram leaves one with a Dynkin diagram of the af\/f\/ine or f\/inite type.
    The hyperbolic KM algebras can be completely classif\/ied
    \cite{Sac,BS} and have  rank $2 \leq r \leq 10$. For $r \geq 3$
    there is a f\/inite number of hyperbolic algebras.  For rank 10, there
    are four algebras, known as $E_{10}$, $BE_{10}$, $CE_{10}$  and $DE_{10}$.
   Hyperbolic KM algebras  appeared in ordinary gravity~\cite{FF} (${\cal F}_3 = AE_3 = H_3$),  supergravity: \cite{J,Miz} ($E_{10}$),
   \cite{Nic1} (${\cal F}_3$), strings \cite{Moore} etc.

   The growth of interest in hyperbolic algebras in theoretical and
   mathematical physics  appeared in 2001
   after the publication of Damour and Henneaux \cite{DamH3}
   devoted to a description of chaotic (BKL-type \cite{BLK})
   behaviour near the singularity  in  string inspired low energy models (e.g.\
   supergravitational ones)   \cite{DamH1} (see also~\cite{DamHJN}).
   It should be noted that these results were based on a billiard approach
   in multidimensional cosmology with dif\/ferent matter sources
   (for $D=4$ suggested by Chitre~\cite{Chitre})   elaborated in
   our papers \cite{IKM1,IKM2,IMb2,IMb2a,IMb3}
   (for a review see also~\cite{DHN,IM-brev}).

  The description of oscillating behaviour near the singularity
  in $D=11$ supergravity~\cite{CJS} (which is related to $M$-theory~\cite{HT,Wit})
  in terms of  motion of a point-like particle in a $9$-dimensional
  billiard  (of f\/inite volume)  corresponding to the Weyl chamber
  of the hyperbolic KM algebra  $E_{10}$ inspired
   another description of  $D=11$ supergravity
   in \cite{DHN-E10}:
    a  formal ``small tension'' expansion of $D=11$ supergravity
  near a space-like singularity was shown to be equi\-valent
  (at least up to 30th order in height) to a null geodesic motion
  in the inf\/inite dimensional coset space ${\cal E}_{10}/K({\cal E}_{10})$
  (here $K({\cal E}_{10})$ is the maximal compact subgroup of the
  hyperbolic Kac--Moody group ${\cal E}_{10}(R)$).

  Recall that $E_{10}$ KM algebra is an over-extension of the f\/inite dimensional
  Lie algebra $E_8$, i.e. $E_{10} = E_8^{++}$. But there is
  another  extension of $E_8$ -- the so-called the very extended
  Kac--Moody algebra of the $E_8$ algebra -- called $E_{11} =
  E^{+++}_8$.
  (To get an understanding of very extended algebras and some of their properties
  see \cite{GOW} and references  therein).
  It has been  proposed by P.~West that the Lorentzian (non-hyperbolic)
  KM algebra $E_{11}$   is responsible for  a hidden algebraic structure
  characterizing  $11$-dimensional supergravity  \cite{W}.
   The same very extended algebra occurs in $IIA$
  \cite{W} and $IIB$ supergravities \cite{SW}. Moreover,
 it was conjectured that an analogous hidden structure is realized in the
 ef\/fective action of the bosonic string (with the KM
 algebra  $k_{27}= D^{+++}_{24}$) \cite{W}  and also for pure $D$ dimensional
 gravity (with the KM algebra~$A^{+++}_{D - 3}$~\cite{LW}).
 It has been suggested in~\cite{EHTW}
 that all the so-called maximally oxidised theories (see also~\cite{HPS}),
 possess the very extension  $G^{+++}$ of the simple Lie algebra $G$.
 It was shown  in~\cite{EHW} that the BPS solutions of  the oxidised
 theory of  a simply laced group ${\cal G}$ form representations of a subgroup
 of the Weyl transformations of the algebra~$G^{+++}$.

   In this paper we brief\/ly review another possibility
   for utilizing  non-singular (e.g.~hyperbolic) KM algebras
   suggested  in three of our papers \cite{IMBl,GrI,IKM}.
   This possibility (implicitly assumed also in
   \cite{IK,IMp1,IMp2,IMp3,Iflux,I-sbr,IM-top})  is related to certain classes of
   exact solutions describing intersecting composite branes
   in a multidimensional gravitational model containing scalar f\/ields
   and antisymmetric forms  def\/ined on (warped) product
   manifolds $M = M_0 \times M_1 \times \cdots \times M_n$, where Einstein factor spaces $M_i$ ($i \geq 1$)
                       are Ricci-f\/lat (at least) for  $i \geq 2$.
      From a pure mathematical point of view these solutions
   may be obtained from sigma-models or Toda chains
   corresponding to certain non-singular KM algebras.
   The information about the (hidden) KM algebra is encoded in
   intersection rules which relate the dimensions of brane
   intersections  with non-diagonal components of the
   generalized Cartan matrix~$A$~\cite{IMJ}.
   We deal here with  generalized Cartan matrices of the
   form
   \begin{gather}
   A_{ss'} \equiv \frac{2(U^s,U^{s'})}{(U^{s'},U^{s'})}, \qquad s,s'\in S,\label{1.1}
   \end{gather}
with $(U^s,U^s)\neq 0$,
   for all $s\in S$ ($S$ is a f\/inite set).
  Here $U^s$ are the so-called brane \mbox{(co-)}vectors.
   They are linear functions on $\R^N$, where $N = n+l$ and
  $l$ is the number of scalar f\/ields. The indef\/inite
  scalar product $(\cdot,\cdot)$ \cite{IMC} is def\/ined on $(\R^N)^{*}$
  and has the signature $(-1,+1, \ldots, +1)$
  when all scalar f\/ields have positive kinetic terms, i.e.\ there
 are no phantoms (or ghosts).   The matrix   $A$ is symmetrizable.
  $U^s$-vectors may be put in one-to-one correspondence with simple roots
  $\alpha_s$ of the generalized KM algebra after
  a suitable normalizing.
  For indecomposable $A$ (when the
  KM algebra is simple) the matrices  $((U^s, U^{s'}))$
  and $((\alpha_s|\alpha_{s'}))$ are proportional to each
  other. Here $(\cdot|\cdot)$ is
  a non-degenerate bilinear symmetric form on a root
  space obeying
  $(\alpha_s|\alpha_{s}) > 0$ for all simple roots $\alpha_{s}$ \cite{Kac}.

  We note that the minisuperspace bilinear form $(\cdot,\cdot)$
  coming from multidimensional gravitational model \cite{IMC} ``does not
  know'' about the def\/inition of $(\cdot|\cdot)$ in \cite{Kac} and hence
  there exist physical examples where all $(U^s,U^s)$
  are negative. Some  examples of this are given below in  Section~\ref{section5}.
   For   $D=11$ supergravity   and ten dimensional  $IIA$, $IIB$
   supergravities all $(U^s,U^s) = 2$ \cite{IMJ,EH}
   and corresponding KM algebras are simply laced.
    It was shown in  our papers \cite{IMb2,IMb2a,IMb3}
    that the inequality $(U^s,U^s)> 0$   is a necessary
   condition for the formation  of the billiard
   wall (in one approaches the singularity) by the $s$-th
   matter source (e.g.\ a f\/luid component or a brane).

   The scalar products  for brane vectors $U^s$  were
 found in \cite{IMC} (for the electric case see also
 \cite{IM11,IM12,IMR})
 \begin{gather}
 \big(U^s,U^{s'}\big)= d_{ss'}+\frac{d_s d_{s'}}{2-D}+
 \chi_s \chi_{s'} \langle \lambda_{s}, \lambda_{s'}\rangle , \label{1.2}
 \end{gather}
where $d_s$ and $d_{s'}$ are dimensions of the brane worldvolumes
corresponding to branes $s$ and $s'$ respectively, $d_{ss'}$ is
the dimension of intersection of the brane worldvolumes, $D$ is
the total space-time dimension, $\chi_s = + 1, -1$ for electric or
magnetic brane respectively, and $\langle \lambda_{s}, \lambda_{s'}\rangle $ is
the non-degenerate scalar product of the $l$-dimensional dilatonic
coupling vectors $\lambda_{s}$ and $\lambda_{s'}$  corresponding
to branes $s$ and $s'$.

 Relations (\ref{1.1}), (\ref{1.2}) def\/ine the brane intersection
 rules \cite{IMJ}
  \[ d_{s s'}= d_{s s'}^{o} + \frac12 K_{s'} A_{s s'}, \]
 $s \ne s'$, where $K_s = (U^s,U^s)$ and
 \begin{gather}
  d_{s s'}^{o} =  \frac{d_s d_{s'}}{D-2} -
    \chi_s\chi_{s'} \langle \lambda_{s}, \lambda_{s'}\rangle \label{1.4}
 \end{gather}
  is the dimension of the so-called
  orthogonal (or ($A_1 \oplus A_1$)-) intersection of
  branes following from the orthogonality condition
  \cite{IMC}
  \begin{gather}
  \big(U^s,U^{s'}\big)= 0, \label{1.5}
  \end{gather}
  $s \ne s'$.
  The orthogonality relations (\ref{1.5}) for brane intersections
  in the non-composite electric case were suggested in \cite{IM11,IM12}
  and for the composite electric   case in~\cite{IMR}.

  Relations (\ref{1.2}) and (\ref{1.4}) were derived in~\cite{IMC}
  for rather general assumptions: the branes were composite, the number of
  scalar f\/ields $l$ was arbitrary as well as the signature of the bilinear form~$\langle \cdot,\cdot\rangle $
  (or, equivalently, the signature of the kinetic term for scalar f\/ields),
  (Einstein) factor spaces $M_i$ had arbitrary dimensions $d_i$ and
  signatures. The intersection rules  (\ref{1.4}) appeared earlier
  (in dif\/ferent notations) in  \cite{AR,AEH,AIR}
  when all $d_i =1$ ($i > 0$) and $\langle\cdot ,\cdot\rangle $ was positive def\/inite
  (in~\cite{AR,AEH}  $l = 1$  and total space-time  had a
  pseudo-Euclidean signature).
    The intersection rules~(\ref{1.4}) were also used in
   \cite{AIV,Oh,BIM,IMJ}  in the context of intersecting
   black brane solutions.

  It  was proved in \cite{Iv3} that the target space
  of the sigma model describing composite electro-magnetic brane
  conf\/igurations on  the product of several Ricci-f\/lat spaces
  is a homogeneous (coset) space $G/H$. It is
  locally symmetric (i.e.\ the Riemann tensor
   is covariantly constant: $\nabla \,{\rm Riem} =0$)
  if and only if
  \[\big(U^{s}-U^{s'}\big)\big(U^{s},U^{s'}\big)=0
  \]
  for all $s$ and $s'$, i.e.\ when any  two brane vectors $U^{s}$ and
  $U^{s'}$, $s \ne s'$, are either coinciding $U^{s} = U^{s'}$
  or orthogonal $(U^{s},U^{s'})=0$
  (for two electric branes and $l=1$ see also~\cite{GR}).

  Now relations for brane vectors $U^s$~(\ref{1.1}) and~(\ref{1.2})
   (with $U^s$ being identif\/ied with roots~$\alpha_s$)
  are widely used in the $G^{+++}$-approach~\cite{EHW,HPS}.
  The orthogonality condition~(\ref{1.5}) describing the
  intersection of branes~\cite{IM11,IM12,IMR,IMC}
  was rediscovered in \cite{EH} (for some particular intersecting
 conf\/igurations of M-theory it was done in \cite{West}).
  It  was found in the context   of $G^{+++}$-algebras
  that the intersection rules for extremal branes are encoded in
  orthogonality conditions between the various roots from which the
  branes arise, i.e.~$(\alpha_s|\alpha_{s'}) = 0$, $s \ne s'$, where
  $\alpha_s$ should be real positive roots
  (``real'' means that $(\alpha_s|\alpha_{s}) > 0$).
  In \cite{EH} another condition on brane, root vectors was found:
  $\alpha_s + \alpha_{s'}$ should not be a root, $s \ne s'$. The last
  condition is trivial  for M-theory and for $IIA$ and $IIB$
  supergravities, but for low energy ef\/fective actions of
  heterotic strings it forbids certain intersections
  that does not take place due to  non-zero
  contributions of Chern--Simons terms.

  It should be noted that the orthogonality relations for
  brane intersections   (\ref{1.5}) which appeared in 1996--97,
  were not well understood by the superstring
  community at that time. The standard intersection rules
  (\ref{1.4})  gave back the well-known zero binding energy conf\/igurations
 preserving some supersymmetries. These brane conf\/igurations were
 originally derived from supersymmetry and duality arguments (see
 for example \cite{Ts-hs,BREJS,Gaunt} and reference therein)
 or by using a no-force condition (suggested for M-branes  in~\cite{Ts-nf}).

\section{The model}\label{section2}

 \subsection{The action}

We consider the model governed by action
 \begin{gather}
 S = \frac{1}{2\kappa^{2}}
 \int_{M} d^{D}z \sqrt{|g|} \Bigg\{ {R}[g] - 2 \Lambda - h_{\alpha\beta}
 g^{MN} \partial_{M} \varphi^\alpha \partial_{N} \varphi^\beta
 \nonumber  \\
 \phantom{S=} - \sum_{a \in {\bf \Delta}}
 \frac{\theta_a}{n_a!} \exp[ 2 \lambda_{a} (\varphi) ] (F^a)^2_g \Bigg\}
 + S_{GH},   \label{2.1}
 \end{gather}
where $g = g_{MN} dz^{M} \otimes dz^{N}$ is the metric on the
manifold $M$, ${\dim M} = D$, $\varphi=(\varphi^\alpha)\in \R^l$
is a~vector of dilatonic scalar f\/ields,
 $(h_{\alpha \beta})$ is a non-degenerate symmetric
 $l\times l$ matrix ($l \in {\mathbb N}  $),
 $\theta_a  \neq 0$,
 \[
  F^a =  dA^a
    =  \frac{1}{n_a!} F^a_{M_1 \ldots M_{n_a}}
       dz^{M_1} \wedge \cdots \wedge dz^{M_{n_a}}
 \]
is an $n_a$-form ($n_a \geq 2$) on a $D$-dimensional manifold $M$,
 $\Lambda$ is a cosmological constant
and $\lambda_{a}$ is a $1$-form on $\R^l$ :
 $\lambda_{a} (\varphi) =\lambda_{a \alpha}\varphi^\alpha$,
 $a \in {\bf \Delta}$, $\alpha=1,\ldots,l$.
In (\ref{2.1}) we denote $|g| = |\det (g_{MN})|$, $(F^a)^2_g =
 F^a_{M_1 \ldots M_{n_a}} F^a_{N_1 \ldots N_{n_a}} g^{M_1 N_1}
 \cdots g^{M_{n_a} N_{n_a}}, $ $a \in {\bf \Delta}$,
 where ${\bf \Delta}$ is
 some f\/inite set (for example, of positive integers),
 and $S_{\rm GH}$ is the standard Gibbons--Hawking boundary
 term \cite{GH}. In models with one time all $\theta_a
  =  1$  when the signature of the metric is $(-1,+1, \ldots, +1)$.
 $\kappa^{2}$ is the multidimensional gravitational constant.

 \subsection{Ansatz for composite  branes}

Let us consider the manifold
 \begin{gather}
 M = M_{0}  \times M_{1} \times \cdots \times M_{n}, \label{2.10}
 \end{gather}
with the metric
 \begin{gather}
 g= e^{2{\gamma}(x)} \hat{g}^0  +
 \sum_{i=1}^{n} e^{2\phi^i(x)} \hat{g}^i, \label{2.11}
 \end{gather}
where $g^0  = g^0 _{\mu \nu}(x) dx^{\mu} \otimes dx^{\nu}$ is an
arbitrary metric with any signature on the manifold $M_{0}$ and
$g^i  = g^{i}_{m_{i} n_{i}}(y_i) dy_i^{m_{i}} \otimes
dy_i^{n_{i}}$ is a metric on $M_{i}$  satisfying the equation
 \begin{gather}
  R_{m_{i}n_{i}}[g^i ] = \xi_{i} g^i_{m_{i}n_{i}}, \label{2.13}
 \end{gather}
 $m_{i},n_{i}=1, \ldots, d_{i}$; $\xi_{i}= {\rm const}$,
 $i=1,\ldots,n$. Here $\hat{g}^{i} = p_{i}^{*} g^{i}$ is the
pullback of the metric $g^{i}$  to the manifold  $M$ by the
canonical projection: $p_{i} : M \rightarrow  M_{i}$,
 $i = 0,\ldots, n$. Thus, $(M_i, g^i )$  are Einstein spaces,
 $i = 1,\ldots, n$.
The functions $\gamma, \phi^{i} : M_0 \rightarrow \R $ are smooth.
We denote $d_{\nu} = {\rm dim} M_{\nu}$; $\nu = 0, \ldots, n$;
 $D = \sum_{\nu = 0}^{n} d_{\nu}$.
We put any manifold $M_{\nu}$, $\nu = 0,\ldots, n$, to be oriented
and connected. Then the volume $d_i$-form
 \[ \tau_i  \equiv \sqrt{|g^i(y_i)|}
  \ dy_i^{1} \wedge \cdots \wedge dy_i^{d_i}, \]
 and signature parameter
\[ \varepsilon(i)  \equiv {\rm sign}( \det (g^i_{m_i n_i})) = \pm 1 \]
 are correctly def\/ined for all $i=1,\ldots,n$.

Let $\Omega = \Omega(n)$  be a set of all non-empty subsets of $\{
1, \ldots,n \}$. The number of elements in $\Omega$ is $|\Omega| =
2^n - 1$. For any $I = \{ i_1, \ldots, i_k \} \in \Omega$, $i_1 <
\ldots < i_k$, we denote
 \begin{gather*}
  \tau(I) \equiv \hat{\tau}_{i_1}
   \wedge \cdots \wedge \hat{\tau}_{i_k}, \\
\eps(I) \equiv \eps(i_1) \cdots \eps(i_k),  \\
 M_{I} \equiv M_{i_1}  \times  \cdots \times M_{i_k}, \\
 d(I) \equiv  \sum_{i \in I} d_i.
\end{gather*}

Here $\hat{\tau}_{i} = p_{i}^{*} \hat{\tau}_{i}$ is the pullback
of the form $\tau_i$  to the manifold  $M$ by the canonical
projection: $p_{i} : M \rightarrow  M_{i}$, $i = 1,\ldots, n$. We
also put $\tau(\varnothing)= \eps(\varnothing)= 1$ and
$d(\varnothing)=0$.

For f\/ields of forms we consider the following composite
electromagnetic ansatz
 \begin{gather}
 F^a=\sum_{I\in\Omega_{a,e}}{\cal F}^{(a,e,I)}+
 \sum_{J\in\Omega_{a,m}}{\cal F}^{(a,m,J)},   \label{2.1.1}
\end{gather}
 where
  \begin{gather}
  {\cal F}^{(a,e,I)}=d\Phi^{(a,e,I)}\wedge\tau(I), \label{2.1.2}\\
  {\cal F}^{(a,m,J)}=
  e^{-2\lambda_a(\varphi)}*(d\Phi^{(a,m,J)} \wedge\tau(J)) \label{2.1.3}
  \end{gather}
 are  elementary forms of electric and magnetic types respectively,
 $a \in {\bf \Delta}$, $I \in \Omega_{a,e}$, $J \in \Omega_{a,m}$ and
 $\Omega_{a,v} \subset \Omega$, $v = e,m$. In (\ref{2.1.3})
 $*=*[g]$ is the Hodge operator on $(M,g)$.

For scalar functions we put
 \begin{gather}
 \varphi^\alpha=\varphi^\alpha(x), \qquad
 \Phi^s=\Phi^s(x), \qquad s\in S.  \label{2.1.5}
 \end{gather}
Thus, $\varphi^{\alpha}$ and $\Phi^s$ are functions on
$M_0$.

Here and below
 \begin{gather} \label{2.1.6} S=S_e \sqcup S_m, \qquad
  S_v= \sqcup_{a \in {\bf \Delta}} \{ a \} \times \{v \} \times \Omega_{a,v},\qquad v=e,m.
  \end{gather}
 Here and in what follows $\sqcup$ means the union of
non-intersecting sets. The set $S$ consists of elements
 $s=(a_s,v_s,I_s)$, where $a_s \in {\bf \Delta}$ is color index,
 $v_s = e, m$ is electro-magnetic index and set $I_s \in
 \Omega_{a_s,v_s}$ describes the location of brane.

Due to (\ref{2.1.2}) and (\ref{2.1.3})
 \begin{gather}
 d(I)=n_a-1, \qquad d(J)=D-n_a-1,  \label{2.1.7}
 \end{gather}
 for  $I \in \Omega_{a,e}$ and $J \in \Omega_{a,m}$
 (i.e.\ in the electric and magnetic case, respectively).

\subsection{The sigma model}\label{section2.3}

Let $d_0 \neq 2$ and
 \[ \gamma=\gamma_0(\phi) \equiv
 \frac1{2-d_0}\sum_{j=1}^nd_j\phi^j, \]
i.e. the generalized harmonic gauge (frame) is used.

Here we put two restrictions on  sets of branes that guarantee the
block-diagonal form of the  energy-momentum tensor and the
existence of the sigma-model representation (without additional
constraints):
\begin{gather}
 {\bf (R1)} \quad d(I \cap J) \leq
 d(I) - 2, \label{2.2.2a}
 \end{gather}
 for any $I,J \in\Omega_{a,v}$, $a \in {\bf \Delta}$, $v= e,m$
(here $d(I) = d(J)$) and
 \begin{gather}
 {\bf (R2)} \quad d(I \cap J)
 \neq 0 \quad \mbox{for} \ \ d_0 = 1, \qquad d(I \cap J) \neq 1 \quad \mbox{for} \ \ d_0 = 3.          \label{2.2.3a}
 \end{gather}

It was proved in \cite{IMC} that equations of motion for the model
(\ref{2.1}) and the Bianchi identities:
  \[   d{\cal F}^s=0,\qquad  s \in S_m, \]
for f\/ields from (\ref{2.11}),
 (\ref{2.1.1})--(\ref{2.1.5}), when Restrictions (\ref{2.2.2a}) and (\ref{2.2.3a})
are  imposed, are equivalent to the equations of motion for the
$\sigma$-model governed by the action
 \begin{gather}
 S_{\sigma 0} = \frac{1}{2 \kappa_0^2}
 \int d^{d_0}x\sqrt{|g^0|}\biggl\{R[g^0]-\hat G_{AB}
 g^{0\mu\nu}\p_\mu\sigma^A\p_\nu\sigma^B \nonumber \\
\phantom{S_{\sigma 0} =}{}  -\sum_{s\in S}\eps_s \exp{(-2U_A^s\sigma^A)}
 g^{0\mu\nu} \p_\mu\Phi^s\p_\nu\Phi^s - 2V \biggr\}, \label{2.2.7}
 \end{gather}
where $(\sigma^A)=(\phi^i,\varphi^\alpha)$, $k_0 \neq 0$, the
index set  $S$ is def\/ined in (\ref{2.1.6}),
 \[ V = {V}(\phi)
   = \Lambda e^{2 {\gamma_0}(\phi)}
    -\frac{1}{2}   \sum_{i =1}^{n} \xi_i d_i e^{-2 \phi^i
    + 2 {\gamma_0}(\phi)} \]
 is the potential,
 \begin{gather}
 (\hat G_{AB})=\barr{cc}
  G_{ij}& 0\\
  0& h_{\alpha\beta}
 \earr       \label{2.2.9}
 \end{gather}
 is the target space metric with
  \[ G_{ij}= d_i \delta_{ij}+\frac{d_i d_j}{d_0-2} \]
  and co-vectors
  \begin{gather}
  U_A^s =
  U_A^s \sigma^A = \sum_{i \in I_s} d_i \phi^i - \chi_s \lambda_{a_s}(\varphi), \!\!\qquad
  (U_A^s) =  (d_i \delta_{iI_s}, -\chi_s \lambda_{a_s \alpha}), \!\! \qquad s=(a_s,v_s,I_s).\!\!\label{2.2.11}
  \end{gather}
Here $\chi_s= +1$ for $v_s = e$ and $\chi_s= -1$ for $v_s = m$;
  \[ \delta_{iI}=\sum_{j\in I}\delta_{ij} \]
 is an indicator of $i$ belonging
to $I$: $\delta_{iI}=1$ for $i\in I$ and $\delta_{iI}=0$ otherwise; and
 \begin{gather}
 \eps_s=(-\eps[g])^{(1-\chi_s)/2}\eps(I_s) \theta_{a_s}, \qquad s\in S, \qquad \eps[g] \equiv {\rm sign}  \det(g_{MN}).\label{2.2.13}
 \end{gather}
More
explicitly (\ref{2.2.13}) reads
  \[
 \eps_s=\eps(I_s) \theta_{a_s} \quad {\rm for} \ \  v_s = e, \qquad
 \eps_s = -\eps[g] \eps(I_s) \theta_{a_s} \quad {\rm for} \  \ v_s = m.
 \]

For f\/inite internal space volumes $V_i$ (e.g.~compact $M_i$) and
electric $p$-branes  (i.e.\ all $\Omega_{a,m} = \varnothing$) the
action (\ref{2.2.7}) coincides with the action (\ref{2.1}) when
$\kappa^{2} = \kappa^{2}_0 \prod_{i=1}^{n} V_i$.

 \section{Solutions governed by harmonic functions}\label{section3}

 \subsection[Solutions with block-orthogonal
             set of $U^s$ and Ricci-flat factor-spaces]{Solutions with block-orthogonal
             set of $\boldsymbol{U^s}$ and Ricci-f\/lat factor-spaces}\label{section3.1}

Here we consider a special class of solutions to equations of
motion governed by several harmonic functions when all factor
spaces are Ricci-f\/lat and the cosmological constant is zero, i.e.\
 $\xi_i = \Lambda = 0$, $i = 1,\ldots,n$. In certain situations
these solutions describe extremal black branes charged by f\/ields
of forms.

The solutions crucially depend upon  scalar products
of $U^s$-vectors $(U^s,U^{s'})$; $s,s' \in S$, where
 \begin{gather}
 (U,U')=\hat G^{AB} U_A U'_B,  \label{3.1.1}
 \end{gather}
for $U = (U_A), U' = (U'_A) \in \R^N$, $N = n + l$ and
 \[
 (\hat G^{AB})=\left(\begin{array}{cc}
 G^{ij}&0\\
 0&h^{\alpha\beta}
 \end{array}\right)
 \]
is the inverse matrix  to  the matrix
 (\ref{2.2.9}). Here as in \cite{IMZ} we have
 \[  G^{ij}=\frac{\delta^{ij}}{d_i}+\frac1{2-D}, \qquad i,j=1,\dots,n.\]

The scalar products (\ref{3.1.1}) for vectors $U^s$  were
calculated in \cite{IMC} and are given by
  \begin{gather}
 (U^s,U^{s'})=d(I_s\cap I_{s'})+\frac{d(I_s)d(I_{s'})}{2-D}+
 \chi_s\chi_{s'}\lambda_{a_s \alpha} \lambda_{a_{s'} \beta} h^{\alpha
 \beta},  \label{3.1.4}
   \end{gather}
where $(h^{\alpha\beta})=(h_{\alpha\beta})^{-1}$, and
 $s=(a_s,v_s,I_s)$, $s'=(a_{s'},v_{s'},I_{s'})$ belong to $S$. This
relation is a~very important one since it encodes  brane data
(e.g.\ intersections) via the  scalar products of $U$-vectors.

Let
  \begin{gather}
 S=S_1 \sqcup \cdots \sqcup S_k, \qquad  S_i\ne\varnothing, \qquad i=1,\dots,k , \label{3.1.5}
  \end{gather}
and
  \begin{gather}
 (U^s,U^{s'})=0                   \label{3.1.6}
  \end{gather}
for all $s\in S_i$, $s'\in S_j$, $i\ne j$; $i,j=1,\dots,k$.
Relation
 (\ref{3.1.5}) means that the set $S$ is a union of $k$ non-intersecting
(non-empty) subsets $S_1,\dots,S_k$. According to (\ref{3.1.6}) the set of
vectors $(U^s, s \in S)$ has a block-orthogonal structure with respect to
the scalar product (\ref{3.1.1}), i.e.\ it  splits into $k$ mutually
orthogonal blocks $(U^s, s \in S_i)$, $i=1,\dots,k$.

Here we consider exact solutions in the model (\ref{2.1}), when
vectors $(U^s,s\in S)$ obey the block-orthogonal decomposition
(\ref{3.1.5}), (\ref{3.1.6}) with scalar products def\/ined in
(\ref{3.1.4}) \cite{IMBl}. These solutions were obtained from the
corresponding solutions to the $\sigma$-model equations of motion~\cite{IMBl}.

\begin{proposition}\label{proposition1}
 Let $(M_0,g^0)$ be Ricci-flat:
 $R_{\mu\nu}[g^0]=0$.
Then the field configuration
 \begin{gather*}
   g^0, \qquad \sigma^A=\sum_{s\in S}\eps_sU^{sA}\nu_s^2\ln H_s, \qquad
   \Phi^s=\frac{\nu_s}{H_s},\qquad   s\in S ,             
 \end{gather*}
satisfies the field equations corresponding to the
action \eqref{2.2.7} with $V=0$ if the real numbers
 $\nu_s$ obey the relations
 \begin{gather}
 \sum_{s'\in S}\big(U^s,U^{s'}\big)\eps_{s'}\nu_{s'}^2 = -1, \qquad s\in S,  \label{3.1.8}
 \end{gather}
 the functions $H_s >0$ are harmonic, i.e.\
 $\Delta[g^0]H_s=0$,
 $s\in S$, and $H_s$ are coinciding inside blocks:
 $H_s=H_{s'}$
for $s,s'\in S_i$, $i=1,\dots,k$.
\end{proposition}

Using the sigma-model solution from Proposition~\ref{proposition1} and the
relations for contra-variant components \cite{IMC}:
 \begin{gather*}
 U^{si}=\delta_{iI_s}-\frac{d(I_s)}{D-2}, \qquad
 U^{s\alpha}=-\chi_s\lambda_{a_s}^\alpha, \qquad s=(a_s,v_s,I_s), 
 \end{gather*}
 we get \cite{IMBl}:
\begin{gather}
 g= \left(\prod_{s\in S}H_s^{2d(I_s)\eps_s\nu_s^2}\right)^{1/(2-D)}
 \left\{\hat{g}^0+ \sum_{i=1}^n
 \left(\prod_{s\in S}H_s^{2\eps_s\nu_s^2\delta_{iI_s}}\right)
 \hat{g}^i\right\},  \label{3.1.11} \\
 \label{3.1.14}
 \varphi^\alpha=-\sum_{s\in S}\lambda_{a_s}^\alpha\chi_s
 \eps_s\nu_s^2\ln H_s, \\
 \label{3.1.15}
 F^a=\sum_{s\in S}{\cal F}^s\delta_{a_s}^a,
 \end{gather}
where $i=1,\dots,n$, $\alpha=1,\dots,l$, $a \in {\bf \Delta}$ and
 \begin{gather}
 {\cal F}^s=\nu_s dH_s^{-1}\wedge\tau(I_s) \quad \mbox{for} \ \ v_s=e, \label{3.1.16} \\
 \label{3.1.17}
 {\cal F}^s=\nu_s (*_0dH_s)\wedge\tau(\bar I_s) \quad  \mbox{for}  \ \ v_s=m,
 \end{gather}
 $H_s$ are harmonic functions on $(M_0,g^0)$ which coincide inside blocks
(i.e.\ $H_s=H_{s'}$ for $s,s'\in S_i$, $i=1,\dots,k$)
 and the relations  (\ref{3.1.8}) on the parameters $\nu_s$ are imposed. Here
 the matrix $((U^s,U^{s'}))$ and parameters $\eps_s$, $s\in S$,
 are def\/ined in (\ref{3.1.4}) and
(\ref{2.2.13}), respectively;
 $\lambda_a^\alpha=  h^{\alpha\beta}\lambda_{a \beta}$,
 $*_0=*[g^0]$ is the Hodge operator  on $(M_0,g^0)$ and
  \[  \bar I = \{1, \dots, n \} \setminus I \]
  is the dual set.
 (In (\ref{3.1.17}) we redef\/ined the sign of $\nu_{s}$-parameter.)

\subsection{Solutions related to non-singular KM algebras}\label{section3.2}

Now we study the  solutions (\ref{3.1.11})--(\ref{3.1.17}) in more
detail and show that some of them may be related to non-singular
 KM algebras. We put
 \[   K_s \equiv (U^s,U^s)\neq 0 \]
 for all $s\in S$ and introduce the  matrix $A=(A_{ss'})$:
 \begin{gather}
 A_{ss'} \equiv \frac{2(U^s,U^{s'})}{(U^{s'},U^{s'})},\qquad s,s'\in S.  \label{3.1.2.2}
 \end{gather}
Here  some ordering in $S$ is assumed.

 Using this def\/inition and (\ref{3.1.4}) we obtain the intersection rules
 \cite{IMJ}
 \begin{gather}
 d(I_{s}\cap I_{s'})=\Delta(s,s')+\frac12 K_{s'} A_{s s'}, \label{3.1.2.3}
 \end{gather}
 where $s \ne s'$, and
 \begin{gather}
 \Delta(s,s') =  \frac{d(I_s)d(I_{s'})}{D-2} -
 \chi_s\chi_{s'}\lambda_{a_s \alpha} \lambda_{a_{s'} \beta}
 h^{\alpha \beta} \label{3.1.D}
 \end{gather}
 def\/ines the so-called ``orthogonal'' intersection rules \cite{IMC} (see
 also \cite{AR,AEH,AIR} for $d_i = 1$).

 In $D = 11$ and $D = 10$ ($IIA$ and $IIB$ )
 supergravity models    all $K_s = 2$ and hence
 the intersection rules (\ref{3.1.2.3}) in this case have a simpler form
 \cite{IMJ}:
  \begin{gather}
  d(I_{s}\cap I_{s'})= \Delta(s,s')+  A_{s s'},  \label{3.1.2.3.sl}
  \end{gather}
  where $s \ne s'$, implying $A_{s s'}  = A_{s' s}$. The
  corresponding KM algebra is simply-laced in this case.

 For $\det A \neq 0$ relation (\ref{3.1.8}) may be rewritten in the
 equivalent form
 \begin{gather}
  - \eps_s\nu_s^2(U^s,U^s)= 2 \sum_{s'\in S} A^{ss'} \equiv b_s,
  \label{3.1.2.5}
 \end{gather}
where $s\in S$ and $(A^{ss'})=A^{-1}$. Thus, equation~(\ref{3.1.8}) may
be resolved in terms of $\nu_s$ for certain $\eps_s=\pm1$, $s\in
S$. We note that due to $(\ref{3.1.6})$ the matrix $A$ has a
block-diagonal structure  and, hence, for any $i$-th block the set
of parameters $(\nu_s,  s \in S_i)$ depends upon the matrix
inverse to the matrix  $(A_{s s'};  s,  s' \in S_i)$.

Now  we consider one-block case when the brane intersections are
related to some  non-singular KM algebras.

{\bf  Finite-dimensional Lie algebras \cite{GrI}.}
Let $A$ be a Cartan matrix of a simple f\/inite-dimensional Lie
algebra. In this case $A_{ss'}\in\{0,-1,-2,-3\}$, $s\ne s'$. The
elements of inverse matrix $A^{-1}$ are positive (see Chapter~7 in~\cite{FS}) and hence we get from (\ref{3.1.2.5}) the same
signature relation  as in the  orthogonal case \cite{IMC}:
 \[ \eps_s\big(U^s,U^s\big) < 0,\qquad s\in S .\]

 When all $(U^s,U^{s}) > 0$ we get
 $\eps_{s} < 0$,  $s \in S$.

Moreover, all $b_s$ are natural numbers:
 \begin{gather}
 b_s = n_s \in {\mathbb N}, \qquad   s \in S.\label{3.1.2.6}
 \end{gather}

The integers $n_s$ coincide with the components of the twice dual
Weyl vector in the basis of simple co-roots (see Chather~3.1.7 in~\cite{FS}).

{\bf  Hyperbolic KM algebras.} Let $A$ be a generalized Cartan matrix corresponding to a simple
hyperbolic KM algebra.

For the  hyperbolic algebras the following relations are satisf\/ied
 \begin{gather}
 \eps_s(U^s,U^s) >0,      \label{3.1.2.20}
 \end{gather}
 since all $b_s$ are negative
 in the hyperbolic case \cite{GOW}:
 \begin{gather}
  b_s  < 0,   \qquad \mbox{where} \quad s\in S.          \label{3.1.2.6b}
 \end{gather}

  For $(U^s,U^{s}) > 0$ we get  $\eps_{s} > 0$,
 $s \in S$. If $\theta_{a_s} > 0$ for all $s \in S$,
then
 \[  \eps(I_s) = 1   \quad {\rm for} \ \ v_s = e, \qquad
 \eps(I_s) = - \eps[g] \quad  {\rm for} \  \ v_s = m.  \]

For a metric $g$ of  pseudo-Euclidean signature all $\eps(I_s) = 1$ and, hence,
all branes are Euclidean or should contain even number of time
directions: $2,4, \ldots$. For $\eps[g] = 1$ only magnetic branes
may be pseudo-Euclidean.

\begin{remark}
The inequalities (\ref{3.1.2.6b}) guarantee the existence of a
principal (real) $so(1,2)$ subalgebra for any hyperbolic Kac--Moody
algebra \cite{NO,GOW}. Similarly the inequalities (\ref{3.1.2.6})
imply the existence of a principal $so(3)$ subalgebra for any
f\/inite dimensional (semi-)simple Lie algebra. It was shown in
\cite{GOW} that this property is not just restricted to hyperbolic
algebras, but holds for a wider class of Lorentzian algebras
obeying the inequalities $b_s \leq 0$ for all $s$.
\end{remark}

\begin{example}\label{example1} {\bf $\boldsymbol{{\cal F}_3 = AE_3}$ algebra \cite{IKM}.}
 Now we  consider an example of the solution corresponding to the
 hyperbolic KM algebra ${\cal F}_3$ with the Cartan matrix
  \begin{gather}
    A=\barr{ccc}
    2 &-2 & 0 \\
    -2 & 2  & -1 \\
    0  & -1 & 2 \earr, \label{3.1.2.21}
     \end{gather}
${\cal F}_3$ contains  ${\bf A_1^{(1)} }$ af\/f\/ine
 subalgebra (it corresponds to the Geroch group) and ${\bf A_2}$
 subalgebra. There exists an example of the solution with the
 $A$-matrix (\ref{3.1.2.21}) for $11$-dimensional model governed by
 the action
\[  S= \int d^{11}z \sqrt{|g|} \left\{R[g] - \frac{1}{4!} \big(F^4\big)^2
     - \frac{1}{4!} \big(F^{4*}\big)^2 \right\}, \]
where ${\rm rank } F^{4} = {\rm rank} F^{4*} = 4$. Here $
 {\bf \Delta} = \{ 4, 4* \}$.  We consider a conf\/iguration with two
magnetic $5$-branes  corresponding to the form $F^4$ and one
electric $2$-brane corresponding to the form  $F^{4*}$. We denote
 $S = \{s_1,s_2,s_3 \}$,
 $a_{s_1} = a_{s_3} = 4$, $a_{s_2} = 4*$ and
 $v_{s_1} = v_{s_3} = m$, $v_{s_2} = e$, where
 $d(I_{s_1}) = d(I_{s_3}) = 6$ and $d(I_{s_2}) = 3$.

The intersection rules (\ref{3.1.2.3}) read
 \[   d(I_{s_1} \cap I_{s_2}) = 0, \qquad
  d(I_{s_2} \cap I_{s_3}) = 1, \qquad
  d(I_{s_1} \cap I_{s_3}) = 4.  \]

For the manifold (\ref{2.10}) we put
 $n= 5$ and $d_1 =2$, $d_2 =4$, $d_3 = d_4 =1$, $d_5 = 2$.
The corresponding brane sets  are the following:
 $I_{s_1} = \{1,2 \}$, $I_{s_2} = \{4,5 \}$, $I_{s_3} = \{2,3,4 \}$.

The  solution reads
 \begin{gather}
  g=H^{-12} \left\{ - dt \otimes dt + H^9 \hat{g}^1 + H^{13} \hat{g}^2
  + H^4 \hat{g}^3  + H^{14} \hat{g}^4  + H^{10} \hat{g}^5  \right\},
    \label{3.1.2.30} \\
  F^4= \frac{dH}{dt} \left\{
 \nu_{s_1} \hat{\tau}_3 \wedge \hat{\tau}_4 \wedge \hat{\tau}_5 +
 \nu_{s_3} \hat{\tau}_1 \wedge \hat{\tau}_5 \right\},
  \nonumber
 \\
 F^{4*} = \frac{dH}{dt} \frac{\nu_{s_2}}{H^2} dt \wedge
 \hat{\tau}_4 \wedge \hat{\tau}_5, \nonumber
 \end{gather}
where  $\nu_{s_1}^2  = \frac{9}{2}$, $\nu_{s_2}^2  = 5$ and
 $\nu_{s_3}^2 = 2$ (see (\ref{3.1.2.5})).

All metrics  $g^i$ are Ricci-f\/lat ($i = 1, \ldots, 5$) and have
Euclidean signatures (this agrees with relations (\ref{3.1.2.20})
and (\ref{2.2.13})), and  $H = ht + h_0 > 0$, where $h, h_0$ are
constants. The metric (\ref{3.1.2.30}) may be also rewritten using
the synchronous time variable $t_s$
  \[ g= - dt_s \otimes dt_s + f^{3/5} \hat{g}^1 +
       f^{-1/5} \hat{g}^2 + f^{8/5} \hat{g}^3  + f^{-2/5} \hat{g}^4  +
       f^{2/5} \hat{g}^5, \]
where $f = 5h t_s = H^{-5} > 0$, $h > 0$ and $t_s > 0$. The metric
describes the power-law ``inf\/lation'' in $D =11$. It is singular for
$t_s \to +0$.

 In the next example we consider a chain of the so-called
 $B_D$-models ($D \geq 11$) suggested in~\cite{IMJ}.
 For $D=11$  the $B_D$-model
 coincides with the truncated (i.e.\ without Chern--Simons term)
 bosonic sector of $D=11$  supergravity \cite{CJS} which is related
 to $M$-theory.  For $D=12$ it coincides with
 truncated $12$-dimensional model from \cite{KKLP}
 which may be  related to $F$-theory  \cite{Vafa}.

 {\bf $\boldsymbol{B_D}$-models.}
 The $B_D$-model has the  action \cite{IMJ}
 \begin{gather}
  S_D=\int
  d^Dz\sqrt{|g|}\biggl\{R[g]+ g^{MN}\p_M\vec\varphi\p_N\vec\varphi-
 \sum_{a=4}^{D-7}\frac1{a!}\exp[2\vec\lambda_a\vec\varphi](F^a)^2\biggr\},
 \label{3.1.2.09}
 \end{gather}
 where
  $\vec\varphi=(\varphi^1,\dots,\varphi^l)\in\R^l$,
  $\vec\lambda_a= (\lambda_{a1},\dots,\lambda_{al})\in\R^l$,
  $l=D-11$, ${\rm rank} F^a=a$, $a=4,\dots,D-7$. Here vectors
  $\vec\lambda_a$ satisfy the relations
  \begin{gather}\label{3.1.2.010}
  \vec\lambda_a\vec\lambda_b=N(a,b)-\frac{(a-1)(b-1)}{D-2} =
  \Lambda_{ab},   \qquad
   N(a,b)=\min(a,b)-3  ,
  \end{gather}
  $a,b=4,\dots,D-7$ and
  $\vec\lambda_{D-7}=-2\vec\lambda_4$. For $D>11$ vectors
  $\vec\lambda_4,\dots,\vec\lambda_{D-8}$ are linearly independent.
  (It may be verif\/ied that  matrix $(\Lambda_{ab})$ is positive def\/inite
   and hence the set of vectors obeying (\ref{3.1.2.010}) does exist.)

 The model (\ref{3.1.2.09}) contains $l$ scalar f\/ields with a
 negative kinetic term (i.e.\
 $h_{\alpha\beta}=-\delta_{\alpha\beta}$ in~(\ref{2.1})) coupled to
 several forms (the number of forms is $(l+1)$) .

 For the brane worldvolumes we have the following dimensions (see
 (\ref{2.1.7}))
 \begin{gather*}
   d(I)=3,\dots,D-8, \qquad I \in \Omega_{a,e}, \\
 d(I)=D-5,\dots,6, \qquad I \in \Omega_{a,m}.
\end{gather*}

Thus, there are $(l+1)$ electric and $(l+1)$ magnetic $p$-branes,
 $p=d(I)-1$. In $B_D$-model all $K_s = 2$.
\end{example}

\begin{example}\label{example2} {\bf $\boldsymbol{H_2(q_1,q_2)}$ algebra \cite{IMBl}.}
  Let
 \[
 A =\barr{cc}
   2& -q_1\\
  -q_2& 2 \earr, \qquad q_1 q_2>4,
  \]

  $q_1,q_2\in {\mathbb N}$. This is the Cartan matrix for the hyperbolic KM
 algebra $H_{2}(q_1,q_2)$ \cite{Kac}.
  From (\ref{3.1.2.5}) we get
 \[   \eps_s\nu_s^2(U^s,U^s)(q_1q_2-4)=2q_s+4, \]
 $s\in\{1,2\}=S$.
An example of the $H_{2}(q,q)$-solution for $B_D$-model  with two
electric $p$-branes  ($p=d_1,d_2$), corresponding to $F^a$ and
$F^b$ f\/ields and intersecting on time manifold, is the following~\cite{IMBl}:
 \begin{gather*}
   g=H^{-2/(q-2)}\hat{g}^0-H^{2/(q-2)}dt\otimes
  dt+\hat{g}^1+\hat{g}^2, \nonumber \\
  F^a=\nu_{1}dH^{-1}\wedge dt\wedge \hat{\tau}_1,
  \nonumber \\   F^b = \nu_{2}dH^{-1}\wedge
  dt\wedge \hat{\tau}_2,  \nonumber \\
    \vec\varphi= \big(\vec\lambda_a+\vec\lambda_b\big)(q -2)^{-1}\ln H,
    \nonumber
  \end{gather*}
where $d_0=3$, $d_1=a - 2$, $a=q+4$, $a < b$, $d_2=b-2$, $D=a+b$,
and $\nu_{1}^2 = \nu_{2}^2 = (q - 2)^{-1}$. The signature
restrictions are : $\eps(1)= \eps(2) = -1$. Thus, the space-time
$(M,g)$ should contain at least three time directions. The minimal
$D$ is 15. For $D=15$ we get $a =7$, $b =8$, $d_1 =5$,
 $d_2= 6$ and $q=3$. (Here we have eliminated  a typo in a sign of
  scalar f\/ields that was  originally in~\cite{IMBl}.)
  \end{example}

\begin{remark}
We note that  af\/f\/ine KM algebras (with $\det A = 0$) do not appear
in the solutions
 (\ref{3.1.11})--(\ref{3.1.17}).
Indeed, any af\/f\/ine Cartan matrix satisfy the
relations
 \[ \sum_{s \in S} a_{s} A_{ss'}=0 \]
  with $a_s > 0$ called Coxeter labels \cite{FS}, $s \in S$.
 This relation makes impossible the existence of the solution to
 equation~(\ref{3.1.8}) (see~(\ref{3.1.2.2})).
\end{remark}

{\bf Generalized Majumdar--Papapetrou  solutions.}
 Now we return to a ``multi-block'' solution
 (\ref{3.1.11})--(\ref{3.1.17}).
 Let $M_0=\R^{d_0}$, $d_0>2$, $g^0=\delta_{\mu\nu}dx^\mu\otimes
 dx^\nu$, $d_1=1$ and $g^1=-dt\otimes dt$. For
   \begin{gather}
   H_s=1+\sum_{b\in X_s}\frac{q_{sb}}{|x-b|^{d_0-2}},
   \label{3.1.3.28}
   \end{gather}
   where $X_s$ is f\/inite
 non-empty subset $X_s \subset M_0$, $s\in S$, all $q_{sb}>0$, and
   $X_s=X_{s'}$, $q_{sb} = q_{s'b}$ for $b \in X_s=X_{s'}$, $s, s'
        \in S_j$, $j =1, \ldots, k$.  The harmonic functions
 (\ref{3.1.3.28}) are def\/ined in domain $M_0 \setminus X$,
  $X=\bigcup_{s\in S}X_s$, and  generate the solutions
 (\ref{3.1.11})--(\ref{3.1.17}).

Denote $S(b)\equiv\{s\in S|\quad b \in X_s\}$, $b \in X$. (In the
one-block case, when $k =1$, all $X_s = X$ and $S(b) = S$.)
 We have a horizon at point $b$ w.r.t.\ time $t$, when  $x\to b\in X$, if
 and only if
\begin{gather*}
 \xi_1(b)\equiv \sum_{s\in S(b)}(-\eps_s) \nu_s^2
   \delta_{1I_s} -\frac1{d_0-2}\ge0.
\end{gather*}
This relation follows just from the requirement of the inf\/inite
propagation time of light to $b \in X$.

 {\bf Majumdar--Papapetrou  solution.}
 Recall that the well-known 4-dimensional
 Majumdar--Papapetrou  (MP) solution \cite{MP} corresponding to
 the Lie algebra ${\bf A_1}$  in our notations reads
\begin{gather*}
    g=H^2 \hat{g}^0 - H^{-2} dt\otimes dt, \\
 F=\nu dH^{-1} \wedge dt,
\end{gather*}
 where $\nu^2 = 2$, $g^0 = \sum_{i=1}^{3} dx^i \otimes dx^i$ and
 $H$ is a harmonic function. We have one electric $0$-brane (point)
 ``attached'' to the time manifold; $d(I_s) =1$, $\eps_s= -1$ and
 $(U^s,U^s) = 1/2$. In this case (e.g.\ for the extremal
 Reissner--Nordstr\"om black hole) we get  $\xi_1(b)=1$,
  $b\in X$. Points $b$ are the points of (regular) horizon.

  For certain examples  of f\/inite-dimensional semisimple Lie
  algebras  (e.g.~for ${\bf A_1} \oplus  \cdots  \oplus  {\bf  A_1}$,
  ${\bf A_2}$  etc.)  the poles  $b$ in $H_s$ correspond to (regular) horizons
  and the solution under consideration describes a collection
  of $k$  blocks  of extremal  charged black branes (in equilibrium)~\cite{IMBl}.

     {\bf Hyperbolic KM algebras.}
      Let us consider a generalized one-block ($k=1$) MP solution corresponding to
      a hyperbolic KM algebra such that
      $(U^s,U^{s}) > 0$ for all  $s \in S$. In this case
      all $\eps_{s} > 0$, $s \in S$, and  hence $\xi_1(b) < 0$.
      Thus, any point $b \in X$ is not a point of the horizon.
      (It may be checked using the analysis carried out in~\cite{IMBl}
      that any non-exceptional point $b$ is a~singular one).
      As a consequence, the generalized MP solution
      corresponding to any  hyperbolic KM algebra
       does not describe a collection
      of  extremal  charged black  branes (in equilibrium)
      when all $(U^s,U^s) > 0$.

 \subsection{Toda-like solutions}\label{section3.3}

 \subsubsection{Toda-like Lagrangian}

 Action (\ref{2.2.7}) may be also written in the form
 \begin{gather}
  S_{\sigma 0} =  \frac{1}{2 \kappa_0^2} \int d^{d_0}x\sqrt{|g^0|}
  \{ R[g^0]-  {\cal G}_{\hat A\hat B}(X)
  g^{0 \mu \nu} \p_\mu X^{\hat A} \p_\nu X^{\hat B}   - 2V \},
   \label{3.2.1n}
  \end{gather}
 where $X = (X^{\hat A})=(\phi^i,\varphi^\alpha,\Phi^s)\in {\ R}^{N}$,
 and the minisupermetric
 $  {\cal G}=
 {\cal G}_{\hat A\hat B}(X)dX^{\hat A}\otimes dX^{\hat B} $
 on the minisuperspace  ${\cal M}={\bf R}^{N}$,  $N = n+l+|S|$
 ($|S|$ is the number of elements in $S$) is def\/ined by the relation
 \begin{gather}
 ({\cal G}_{\hat A\hat B}(X))=\left(\begin{array}{ccc}
 G_{ij}&0&0\\[5pt]
 0&h_{\alpha\beta}&0\\[5pt]
 0&0&\eps_s \exp(-2U^s(\sigma))\delta_{ss'}
 \end{array}\right).
 \label{3.2.3n}
 \end{gather}

Here we consider exact solutions to f\/ield equations corresponding
to the action  (\ref{3.2.1n})
 \begin{gather}
 R_{\mu\nu}[g^0]=
 {\cal G}_{\hat A \hat B}(X) \p_{\mu} X^{\hat A} \p_\nu  X^{\hat B}
 + \frac{2V}{d_0-2}g_{\mu\nu}^0, \label{3.2.4}  \\
  \frac{1}{\sqrt{|g^0|}} \p_\mu [\sqrt{|g^0|} {\cal
 G}_{\hat C \hat B}(X)g^{0\mu\nu }  \p_\nu  X^{\hat B}] -
 \frac{1}{2} {\cal G}_{\hat A \hat B, \hat C}(X) g^{0,\mu \nu}
 \p_{\mu} X^{\hat A} \p_\nu  X^{\hat B} = V_{,\hat C},
  \label{3.2.5}
 \end{gather}
 where $s\in  S$. Here  $V_{,\hat C} = \p V / \p  X^{\hat C}$.

We put
 \[   X^{\hat A}(x) =  F^{\hat A}(H(x)), \]
where $F: (u_{-}, u_{+}) \rightarrow \R^{N}$  is a smooth function,
 $H: M_0 \rightarrow \R $ is a harmonic function on $M_0$ (i.e.
 $\Delta[g^0]H=0$), satisfying  $u_{-} < H(x)  < u_{+}$ for all $x \in M_0$.
We take all factor spaces as  Ricci-f\/lat and the cosmological
constant is set to zero, i.e. the relations  $\xi_i = 0$ and
$\Lambda = 0$ are satisf\/ied.

In this case the potential is zero : $V = 0$. It may be verif\/ied
that the f\/ield equations (\ref{3.2.4}) and (\ref{3.2.5}) are
satisf\/ied identically  if $F = F(u)$ obeys the Lagrange equations
for the Lagrangian
 \begin{gather}
  L =  \frac{1}{2} {\cal G}_{\hat A\hat B}(F) \dot F^{\hat A}
  \dot F^{\hat B}  \label{3.2.12}
  \end{gather}
with the zero-energy constraint
 \begin{gather}
  E =  \frac{1}{2} {\cal G}_{\hat A\hat B}(F) \dot F^{\hat A}
 \dot F^{\hat B} = 0. \label{3.2.13}
 \end{gather}
This means that  $F: (u_{-}, u_{+}) \rightarrow  \R^N$ is a
null-geodesic map for the minisupermetric ${\cal G}$. Thus, we are
led to the Lagrange system (\ref{3.2.12}) with the minisupermetric
${\cal G}$ def\/ined in (\ref{3.2.3n}).

The problem of integrability will be simplif\/ied if we integrate
the Lagrange equations corresponding to $\Phi^s$ (i.e. the
Maxwell-type equations for $s\in S_e$ and Bianchi identities for
$s\in S_m$):
 \begin{gather}
 \frac d{du}\left(\exp(-2U^s(\sigma))\dot\Phi^s\right)=0
 \Longleftrightarrow
 \dot\Phi^s=Q_s \exp(2U^s(\sigma)), \label{3.2.14}
 \end{gather}
where $Q_s$ are constants, and $s \in S$. Here $(F^{\hat A})=
(\sigma^A, \Phi^s)$. We put $Q_s\ne0$ for all  $s \in S$.

For f\/ixed $Q=(Q_s,s\in S)$ the Lagrange equations for the Lagrangian
(\ref{3.2.12})  corresponding to $(\sigma^A)=(\phi^i,\varphi^\alpha)$,
when equations (\ref{3.2.14}) are substituted, are equivalent to the
Lagrange equations for the Lagrangian
 \begin{gather}
 L_Q=\frac12\hat G_{AB}\dot \sigma^A\dot \sigma^B-V_Q,
 \label{3.2.16}
 \end{gather}
where
 \begin{gather}
 V_Q=\frac12  \sum_{s\in S}  \eps_s Q_s^2 \exp[2U^s(\sigma)],
 \label{3.2.17}
 \end{gather}
the matrix $(\hat G_{AB})$ is def\/ined in (\ref{2.2.9}). The
zero-energy constraint (\ref{3.2.13}) reads
 \begin{gather}
 E_Q= \frac12 \hat G_{AB}\dot \sigma^A \dot \sigma^B+ V_Q =0.
 \label{3.2.18}
 \end{gather}

\subsubsection{The solutions}\label{section3.3.2}

Here, as above we are interested in exact solutions for a special
case when  $K_s =(U^s,U^s)\neq 0$, for all $s\in S$, and the
generalized Cartan matrix (\ref{3.1.2.2}) is  non-degenerate.
 It follows from the
 non-degeneracy of the matrix (\ref{3.1.2.2}) that
 vectors $U^s$, $s \in S,$ are  linearly independent. Hence, the
 number of  vectors  $U^s$ should not exceed the dimension of
 $\R^{n+ l}$, i.e.\  $|S| \leq n+ l$.

The  exact solutions  were  obtained in \cite{IK} and are
 \begin{gather}
 g= \left(\prod_{s \in S} f_s^{2d(I_s)h_s/(D-2)}\right)
 \Biggl\{ \exp(2c^0 H+ 2\bar  c^0) \hat{g}^0
    \nonumber  \\ \label{3.2.63}
\phantom{g=} + \sum_{i =1}^{n} \left(\prod_{s \in S }
 f_s^{- 2h_s \delta_{i I_s} }\right)
 \exp(2c^i H+ 2\bar  c^i) \hat{g}^i \Biggr\},
 \\  \label{3.2.64}
 \exp(\varphi^\alpha) =
 \left( \prod_{s\in S} f_s^{h_s \chi_s\lambda_{a_s}^\alpha} \right)
 \exp(c^\alpha H +\bar c^\alpha),\qquad \alpha=1,\dots,l,
 \end{gather}
  and $F^a=\sum_{s\in S}{\cal F}^s\delta_{a_s}^a$
 with
 \begin{gather*}
 {\cal F}^s= Q_s
 \left( \prod_{s' \in S}  f_{s'}^{- A_{s s'}} \right) dH \wedge\tau(I_s),
 \qquad s\in S_e, \\ 
 {\cal F}^s
 =  Q_s (*_0 d H) \wedge \tau(\bar I_s), \qquad s\in S_m,
 \end{gather*}
where  $*_0 = *[g^0]$ is the Hodge operator on $(M_0,g^0)$. Here
 \begin{gather}
 f_s = f_s(H) = \exp(- q^s(H)), \label{3.2.65}
 \end{gather}
where $q^s(u)$ is a solution to the  Toda-like equations
 \begin{gather}
 \ddot{q}^s = - B_s \exp\left( \sum_{s' \in S} A_{s s'}  q^{s'} \right)
 \label{3.2.35}
 \end{gather}
with $ B_s = K_s \eps_s Q_s^2$, $s \in S$, and $H = H(x)$ ($x \in
 M_0$) is a harmonic function on $(M_0,g^0)$. Vectors $c=(c^A)$ and
 $\bar c=(\bar c^A)$ satisfy the linear constraints
 \begin{gather}
 U^s(c)=
 \sum_{i \in I_s}d_ic^i-\chi_s\lambda_{a_s\alpha}c^\alpha=0,
 \qquad
 U^s(\bar c)= 0, \qquad s\in S, \label{3.2.47}
 \end{gather}
and
 \begin{gather}
  c^0 = \frac1{2-d_0}\sum_{j=1}^n d_j c^j, \qquad  \bar
  c^0 = \frac1{2-d_0}\sum_{j=1}^n d_j \bar c^j. \label{3.2.52a}
 \end{gather}
The zero-energy constraint reads
 \begin{gather}
  2E_{T} + h_{\alpha\beta}c^\alpha c^\beta+
 \sum_{i=1}^n d_i(c^i)^2+
 \frac1{d_0-2}\left(\sum_{i=1}^nd_ic^i\right)^2 = 0,
  \label{3.2.53}
 \end{gather}
 where
 \begin{gather}
 E_{T} = \frac{1}{4}  \sum_{s,s' \in S} h_s A_{s s'}
 \dot{q^s} \dot{q^{s'}}
  + \sum_{s \in S} A_s  \exp\left( \sum_{s' \in S} A_{s s'} q^{s'} \right)
 \label{3.2.54a}
 \end{gather}
 is an integration constant (energy) for the solutions from
(\ref{3.2.35}) and $A_s =  \frac{1}{2}  \eps_s Q_s^2$.

We note that  equations (\ref{3.2.35}) correspond to the
Lagrangian
\[  L_{T} = \frac{1}{4}  \sum_{s,s' \in S} h_s  A_{s s'}
 \dot{q^s} \dot{q^{s'}} -  \sum_{s \in S} A_s  \exp\left( \sum_{s' \in
 S} A_{s s'} q^{s'} \right), \]
where $h_s = K_s^{-1}$.

Thus, the solution is given by relations
(\ref{3.2.63})--(\ref{3.2.65}) with the functions  $q^s$ being
def\/ined in~(\ref{3.2.35}) and with  relations on the parameters of
solutions $c^A$, $\bar c^A$ $(A= i,\alpha,0)$, imposed by
(\ref{3.2.47}), (\ref{3.2.52a}), (\ref{3.2.53}).

\section{Cosmological-type solutions}\label{section4}

 Now we consider the case $d_0 =1$, $M_0 = \R$, i.e.\
we are interested in applications to the sector with dependence on
 a single variable.
We consider the manifold
 \[  M = (u_{-},u_{+})  \times M_{1} \times \cdots \times M_{n}  \]
with a metric
 \[  g= w e^{2{\gamma}(u)} du \otimes du +
 \sum_{i=1}^{n} e^{2\phi^i(u)} {\hat g}^i, \]
 where $w=\pm 1$, $u$ is a distinguished coordinate which, by
convention, will be called ``time'';
 $(M_i,g^i)$ are oriented and connected Einstein spaces
(see (\ref{2.13})), $i=1,\dots,n$.
The functions $\gamma,\phi^i$: $(u_-,u_+)\to \R$ are smooth.

Here we adopt the brane ansatz from Section~\ref{section2} putting $g^0= w  du
  \otimes du$.

\subsection{Lagrange dynamics}

It follows from Subsection~\ref{section2.3} that the equations of motion  and the
Bianchi identities for the f\/ield conf\/iguration under consideration
(with the restrictions from Subsection~\ref{section2.3} imposed) are equivalent
to equations of motion for 1-dimensional $\sigma$-model with the
action
 \begin{gather}
 S_{\sigma} = \frac{\mu}2
 \int du {\cal N} \biggl\{G_{ij}\dot\phi^i\dot\phi^j
 +h_{\alpha\beta}\dot\varphi^{\alpha}\dot\varphi^{\beta}
 +\sum_{s\in S}\eps_s\exp[-2U^s(\phi,\varphi)](\dot\Phi^s)^2
 -2{\cal N}^{-2}V_{w}(\phi)\biggr\},\!\!\!  \label{4.1.1}
 \end{gather}
where $\dot x\equiv dx/du$,
 \begin{gather}
 V_{w} = -w V = -w\Lambda e^{2\gamma_0(\phi)}+
 \frac{w}{2} \sum_{i =1}^{n} \xi_i d_i e^{-2 \phi^i + 2 {\gamma_0}(\phi)}
  \label{4.1.2}
 \end{gather}
is the potential with $\gamma_0(\phi)
  \equiv\sum_{i=1}^nd_i\phi^i$, and
  ${\cal N}=\exp(\gamma_0-\gamma)>0$ is the modif\/ied lapse function, $U^s =
U^s(\phi,\varphi)$ are def\/ined in (\ref{2.2.11}), $\eps_s$ are
def\/ined in (\ref{2.2.13}) for $s=(a_s,v_s,I_s)\in S$, and
  $G_{ij}=d_i\delta_{ij}-d_id_j$  are components of  ``pure
cosmological'' minisupermetric, $i,j=1,\dots,n$~\cite{IMZ}.

In the electric case $({\cal F}^{(a,m,I)}=0)$ for f\/inite internal space
volumes $V_i$ the action (\ref{4.1.1}) coincides with the
action~(\ref{2.1}) if
  $\mu=-w/\kappa_0^2$, $\kappa^{2} = \kappa^{2}_0 V_1 \cdots V_n$.

Action (\ref{4.1.1}) may be also written in the form
 \begin{gather}
 S_\sigma=\frac\mu2\int du{\cal N}\left\{
 {\cal G}_{\hat A\hat B}(X)\dot X^{\hat A}\dot X^{\hat B}-
 2{\cal N}^{-2}V_w \right\},     \label{4.1.6}
 \end{gather}
where $X = (X^{\hat A})=(\phi^i,\varphi^\alpha,\Phi^s)\in
 {\R}^{N}$, $N = n +l + |S|$, and minisupermetric
 ${\cal G}$ is def\/ined in~(\ref{3.2.3n}).

{\bf Scalar products.}
The minisuperspace metric (\ref{3.2.3n}) may be also written in the form
${\cal G}=\hat G+\sum_{s\in S} \eps_s
          e^{-2U^s(\sigma)}d\Phi^s\otimes d\Phi^s$,
where $\sigma = (\sigma^A) = (\phi^i,\varphi^\alpha)$,
 \[  \hat G=\hat G_{AB}d \sigma^A
 \otimes d \sigma^B=G_{ij}d\phi^i\otimes d\phi^j+
 h_{\alpha\beta}d\varphi^\alpha\otimes d\varphi^\beta, \]
is the truncated minisupermetric and $U^s(\sigma)=U_A^s
 \sigma^A$ is def\/ined in (\ref{2.2.11}). The potential
(\ref{4.1.2}) now reads
  \[  V_w=(-w\Lambda)e^{2U^\Lambda(\sigma)}+\sum_{j=1}^n \frac{w}{2} \xi_jd_j
  e^{2U^j(\sigma)}, \]
  where
  \begin{gather}
   U^j(\sigma)=U_A^j
 \sigma^A=-\phi^j+\gamma_0(\phi), \qquad
 (U_A^j)=(-\delta_i^j+d_i,0), \label{4.1.10}
 \\ \nonumber
 U^\Lambda(\sigma)=U_A^\Lambda \sigma^A=\gamma_0(\phi),
 \qquad (U_A^\Lambda)=(d_i,0).
  \end{gather}

The integrability of the Lagrange system (\ref{4.1.6}) crucially
depends upon the scalar products of co-vectors $U^\Lambda$, $U^j$,
$U^s$ (see~(\ref{3.1.1})). These products are def\/ined by
(\ref{3.1.4}) and the following relations \cite{IMC}
 \begin{gather}
 \big(U^i,U^j\big)=\frac{\delta_{ij}}{d_j}-1, \label{4.1.14}
 \\
 \big(U^i,U^\Lambda\big)=-1, \nonumber
 \qquad
 \big(U^\Lambda,U^\Lambda\big)=-\frac{D-1}{D-2},
 \\
 \big(U^s,U^i\big)=-\delta_{iI_s}, \qquad
 \big(U^s,U^\Lambda\big)=\frac{d(I_s)}{2-D}, \label{4.1.17}
  \end{gather}
 where $s=(a_s,v_s,I_s) \in S$;
 $i,j= 1,\dots,n$.

{\bf Toda-like representation.}
We put $\gamma= \gamma_0(\phi)$, i.e. the harmonic
time gauge is considered.  Integrating
the Lagrange equations corresponding to $\Phi^s$
(see (\ref{3.2.14})) we are led to the Lagrangian
from (\ref{3.2.16}) and the zero-energy constraint (\ref{3.2.18})
with the modif\/ied potential
 \begin{gather}
  V_Q=V_w +\frac12\sum_{s\in S} \eps_sQ_s^2\exp[2U^s(\sigma)],
  \label{4.1.19}
 \end{gather}
where $V_w$ is def\/ined in  (\ref{4.1.2}).

\subsection[Solutions with $\Lambda = 0$]{Solutions with $\boldsymbol{\Lambda = 0}$}

Here we consider  solutions with $\Lambda = 0$.

 \subsubsection{Solutions with Ricci-f\/lat factor-spaces}\label{section4.2.1}

Let all spaces be Ricci-f\/lat, i.e.\ $ \xi_1 =\dots=\xi_n=0$.

Since  $H(u)= u$ is a harmonic function on $(M_0,g^0)$
with $g^0 = w du\otimes du$  we get
for the metric and scalar f\/ields from
(\ref{3.2.63}), (\ref{3.2.64})~\cite{IK}
  \begin{gather}
 g= \left(\prod_{s \in S} f_s^{2d(I_s)h_s/(D-2)}\right)
    \Biggl\{ \exp(2c^0 u+ 2\bar  c^0) w du \otimes du
     \nonumber \\ \label{4.3.3}
\phantom{g=} + \sum_{i =1}^{n} \left(\prod_{s\in S }
 f_s^{- 2h_s \delta_{i I_s} }\right)
 \exp(2c^i u + 2\bar  c^i) {\hat g}^i \Biggr\},
 \\  \label{4.3.4}
 \exp(\varphi^\alpha) = \left( \prod_{s\in S}
 f_s^{h_s \chi_s\lambda_{a_s}^\alpha} \right)
 \exp(c^\alpha u +\bar c^\alpha), \qquad \alpha=1,\dots,l,
  \end{gather}
 and
  $F^a= \sum_{s \in S} \delta^a_{a_s} {\cal F}^{s}$
  with
 \begin{gather}
 {\cal F}^s= Q_s
 \left( \prod_{s' \in S}  f_{s'}^{- A_{s s'}} \right) du \wedge\tau(I_s),
 \qquad s\in S_e, \label{4.3.19c}
  \\
 \label{4.3.19d}
 {\cal F}^s= Q_s \tau(\bar I_s), \qquad s \in S_m,
  \end{gather}
 $Q_s \neq 0$, $s \in S$.
 Here   $f_s = f_s(u) = \exp(- q^s(u))$ and
 $q^s(u)$ obey  Toda-like equations (\ref{3.2.35}).

Relations (\ref{3.2.52a}) and  (\ref{3.2.53}) take the form
 \begin{gather}
  c^0 = \sum_{j=1}^n d_j c^j,
 \qquad  \bar  c^0 = \sum_{j=1}^n d_j \bar c^j,
   \nonumber  \\
  2E_{T} + h_{\alpha\beta}c^\alpha c^\beta+ \sum_{i=1}^n d_i(c^i)^2
  - \left(\sum_{i=1}^nd_ic^i\right)^2 = 0,
  \nonumber
 \end{gather}
 with $E_{T}$  from (\ref{3.2.54a}) and all other relations
(e.g.\ constraints~(\ref{3.2.47})) are unchanged.

 This solution in the special case
 of an ${\bf A_m}$ Toda chain, was obtained earlier  in \cite{GM1} (see
 also~\cite{GM2}). Some special conf\/igurations were considered
 earlier in \cite{LMPX,LPX,LMMP}.

 Currently, the cosmological solutions with branes
 are  considered often in a context of
 $S$-brane terminology  \cite{S1}.  $S$-branes
 were originally space-like analogues of  $D$-branes,
 see also \cite{S2,S3,S4,S5,I-sbr,Ohta,Iohta} and references
 therein.

 \subsubsection{Solutions with one curved factor-space}

The cosmological solution with Ricci-f\/lat spaces
may be also  modif\/ied to the following case:
 $ \xi_1 \ne0$, $\xi_2=\cdots=\xi_n=0$,
i.e. one space is curved and others are Ricci-f\/lat and
 $1 \notin I_s$,  $s  \in S$,
i.e.\ all ``brane'' submanifolds  do not  contain $M_1$.

 The potential (\ref{3.2.17}) is modif\/ied for $\xi_1 \ne0$ as
 follows (see (\ref{4.1.19}))
 \[   V_Q=\frac12  \sum_{s\in S}  \eps_s Q_s^2 \exp[2U^s(\sigma)]
 + \frac12 w\xi_1 d_1  \exp[2U^1(\sigma)], \]
 where $U^1(\sigma)$ is def\/ined in  (\ref{4.1.10}) ($d_1 > 1$).

For the scalar products we get from (\ref{4.1.14}) and (\ref{4.1.17}){\samepage
  \begin{gather}
  (U^1,U^1)=\frac1{d_1}-1<0, \qquad (U^1,U^{s})=0
   \label{4.3.7}
  \end{gather}
for all $s\in S$.}

The solution in the case under consideration may be obtained   by
a little modif\/ication of the solution from the previous section
(using (\ref{4.3.7}), relations $U^{1i}= - \delta_1^i/d_1$,
$U^{1\alpha}=0$) \cite{IK}
 \begin{gather}
 g= \left(\prod_{s \in S} [f_s(u)]^{2 d(I_s)
  h_s/(D-2)} \right) \Biggl\{[f_1(u)]^{2d_1/(1-d_1)}\exp(2c^1u + 2
  \bar c^1) \nonumber  \\
 \phantom{g=}{}  \times[w du \otimes du+ f_1^2(u) \hat{g}^1] +
  \sum_{i = 2}^{n} \left(\prod_{s\in S} [f_s(u)]^{- 2 h_s  \delta_{i
  I_s} } \right)\exp(2c^i u+ 2 \bar c^i) \hat{g}^i \Biggr\}, \label{4.3.19}
 \\  \exp(\varphi^\alpha) =
 \left( \prod_{s\in S} f_s^{h_s \chi_s \lambda_{a_s}^\alpha} \right)
 \exp(c^\alpha u + \bar c^\alpha),  \label{4.3.19a} \\
 F^a= \sum_{s \in S} \delta^a_{a_s} {\cal F}^{s},
 \label{4.3.19b}
  \end{gather}
 with forms  ${\cal F}^{s}$ def\/ined in  (\ref{4.3.19c})
 and (\ref{4.3.19d}).

Here  $f_s = f_s(u) = \exp(- q^s(u))$ where $q^s(u)$ obey
Toda-like equations (\ref{3.2.35}) and{\samepage
 \begin{gather}
 f_1(u) =\left\{\begin{array}{ll}
 R \sinh(\sqrt{C_1}(u-u_1)), &  C_1>0, \ \xi_1
  w > 0; \nonumber  \\
 R \sin(\sqrt{|C_1|}(u-u_1)), &  C_1<0, \  \xi_1 w>0; \nonumber  \\
 R \cosh(\sqrt{C_1}(u-u_1)),  & C_1>0, \ \xi_1w <0; \nonumber \\
 \left|\xi_1(d_1-1)\right|^{1/2}, & C_1=0,  \ \xi_1w>0, \nonumber
\end{array}\right.
 \end{gather}
 $u_1$, $C_1$ are constants and $R =  |\xi_1(d_1-1)/C_1|^{1/2}$.}

 The vectors $c=(c^A)$ and $\bar c=(\bar c^A)$ satisfy the linear constraints
 \[  U^r(c)= U^r(\bar c)= 0, \qquad r = s,1, \]
(for $r =s$ see (\ref{3.2.47})) and the zero-energy constraint
 \[  C_1\frac{d_1}{d_1-1}= 2 E_{T} +
  h_{\alpha\beta}c^\alpha c^\beta+ \sum_{i=2}^nd_i(c^i)^2+
 \frac1{d_1-1}\left(\sum_{i=2}^nd_ic^i\right)^2.  \]

\subsubsection[Special solutions for block-orthogonal set of vectors $U^s$]{Special solutions for block-orthogonal set of vectors $\boldsymbol{U^s}$}
\label{section4.2.3}

Let  us consider block-orthogonal case: (\ref{3.1.5}),
(\ref{3.1.6}). In this case  we get
 \[ f_s = (\bar{f}_s)^{b_s} \]
   where
 $b_s = 2 \sum_{s' \in S} A^{s s'}$, $(A^{ss'})= (A_{ss'})^{-1}$
and
 \begin{gather*}
  \bar{f}_s(u)=\left\{\begin{array}{ll}
 R_s \sinh(\sqrt{C_s}(u-u_s)), &
 C_s>0, \  \eta_s\eps_s<0; \nonumber \\
  R_s \sin(\sqrt{|C_s|}(u-u_s)), &
 C_s<0, \; \eta_s\eps_s<0; \nonumber  \\
  R_s \cosh(\sqrt{C_s}(u-u_s)), &
 C_s>0, \; \eta_s\eps_s>0; \nonumber \\
  \frac{|Q^s|}{|\nu_s|}(u-u_s), &   C_s=0, \  \eta_s\eps_s<0,
\end{array}\right.
 \end{gather*}
where
 $R_s = |Q_s|/(|\nu_s||C_s|^{1/2})$,
  \[   \eta_s \nu_s^2 = b_s h_s, \]
  $\eta_s= \pm 1$, $C_s$, $u_s$ are constants, $s \in S$. The
constants $C_s$, $u_s$ are coinciding inside the blocks:
 $u_s =   u_{s'}$, $C_s = C_{s'}$, $s,s' \in S_i$, $i = 1, \ldots, k$.  The
ratios $\eps_s Q_s^2/(b_s h_s)$  also coincide  inside the blocks,
or, equivalently,
  \begin{gather}
  \frac{\eps_s Q_s^2}{b_s h_s} =
  \frac{\eps_{s'} Q_{s'}^2}{b_{s'} h_{s'}}, \qquad s,s' \in S_i, \quad i = 1, \ldots, k.
  \label{4.3.25n}
  \end{gather}

For energy integration constant (\ref{3.2.54a}) we get
  \begin{gather}
  E_{T} = \frac{1}{2} \sum_{s \in S} C_s b_s h_s.
  \label{4.3.20e}
  \end{gather}

The solution (\ref{4.3.19})--(\ref{4.3.19b}) with a
block-orthogonal set of $U^s$-vectors  was obtained in
\cite{IMJ1,IMJ2} (for non-composite case see also earlier paper by
K.A.~Bronnikov~\cite{Br1}). The generalized KM algebra
corresponding to the generalized Cartan matrix $A$ in this case is
semisimple. In the special orthogonal (or $A_1 \oplus \cdots \oplus
A_1 $) case when: $|S_1| = \cdots = |S_k| = 1$, the solution  was
obtained in \cite{IMJ}.

Thus, here we presented a large class of  exact solutions for
invertible generalized Cartan matrices (e.g.\ corresponding to
hyperbolic KM algebras). These solutions are governed by Toda-type
equations. They are integrable in quadratures for
f\/inite-dimensional semisimple Lie algebras~\cite{T,B,K,OP,AM} in
agreement with  Adler--van Moerbeke criterion \cite{AM} (see also~\cite{KTr}).

The problem of integrability of Toda-chains related to Lorentzian
(e.g.\ hyperbolic) KM algebras is much more complicated than in the
Euclidean case. This is supported by the result from \cite{EmTs}
(based on calculation of the Kovalevskaya exponents) where it was
shown  that the known cases of algebraic integrability for
Euclidean Toda chains have no direct analogues in the case of
spaces with pseudo-Euclidean metrics because the full-parameter
expansions of the general solution contain complex powers of the
independent variable. A similar result, using the Painlev\'e
property, was obtained earlier for 2-dimensional Toda chains
related to hyperbolic KM algebras~\cite{GIMiz}.

\begin{remark}
It was shown in   \cite{FrS}  that all supergravity billiards
corresponding to sigma-models on any $U/H$ non compact-symmetric
space and obtained by compactifying supergravity to $D=3$ are
fully integrable. As far as we know this result could not be
reformulated in terms of integrability of Toda-chains
corresponding to certain Lorentzian (e.g.\ hyperbolic) KM algebras.
\end{remark}

  \subsection[Examples of $S$-brane solutions]{Examples of $\boldsymbol{S}$-brane solutions}

 \begin{example}\label{example3}
 {\bf $\boldsymbol{S}$-brane solution governed by $\boldsymbol{E_{10}}$ Toda chain}

  Let us consider the $B_{16}$-model
  in  $16$-dimensional pseudo-Euclidean space of
  signature $(-, +$, $\dots, +)$
   with six forms $F^4, \dots, F^9$ and f\/ive
  scalar f\/ields $\varphi^1, \dots, \varphi^5$, see~(\ref{3.1.2.09}).
  Recall that for two branes corresponding
  to $F^a$ and $F^b$ forms the orthogonal (or $(A_1 + A_1)$-)
  intersection   rules read \cite{IMJ,IM-top}:
   \begin{gather*}
  (a-1)_e \cap_{o} (b -1)_e = N(a,b) = {\rm min} (a,b) - 3,
 \\
   (a-1)_e \cap_{o} (D- b -1)_m = a - 1 - N(a,b),
   \end{gather*}
  where  $d_{v} \cap_{o} d'_{v'}$  denotes the dimension of
  orthogonal intersection for two branes with the dimensions of
   their worldvolumes being $d$ and $d'$.
   $d_{v} \cap_{o} d'_{v'}$  coincides with the symbol $\Delta(s,s')$
   from (\ref{3.1.D})\footnote{Here as in \cite{IM-top}
   our notations dif\/fer from those adopted in string theory.
   For example for intersection of M2- and M5-branes   we write
    $3 \cap_{o} 6$ = 2 instead of $2 \cap 5$ = 1.}.
  The subscripts $v, v' = e,m$ here indicate whether
  the brane is electric ($e$) or magnetic ($m$) one.
  In what  follows we will be interested in the following
  orthogonal intersections:
   $4_e \cap_{o} 4_e = 2$, $4_e \cap_{o} 5_e = 2$,
   $4_e \cap_{o} 11_m = 3$, $5_e \cap_{o} 11_m = 4$.

  Here we deal with 10 ($S$-)branes: eight electric branes
  $s_1$, $s_2$, $s_3$, $s_4$, $s_5$, $s_6$, $s_8$, $s_9$ corresponding to
   5-form $F^5$, one electric brane
  $s_7$ corresponding to  6-form $F^6$ and
  one magnetic brane $s_{10}$ corresponding to
  4-form $F^4$. The brane sets are as follows:
  $I_1 = \{ 3, 4, 10, 12 \}$,
  $I_2 = \{ 1, 6, 7, 12 \}$,
  $I_3 = \{ 8, 9, 10, 12 \}$,
  $I_4 = \{ 1, 2, 3, 12 \}$,
  $I_5 = \{ 5, 6, 10, 12 \}$,
  $I_6 = \{ 1, 4, 8, 12 \}$,
  $I_7 = \{ 2, 7, 10, 12, 13 \}$,
  $I_8 = \{ 3, 6,  8, 12 \}$,
  $I_9 = \{ 1, 10, 11, 12 \}$,
  $I_{10} = \{ 1, 2, 3, 4, 5, 6, 7, 8, 9, 10, 11 \}$.

 It may be verif\/ied that these sets do obey
 $E_{10}$ intersection rules following from the
  relations (\ref{3.1.2.3.sl}) (with $I_{s_i} = I_i$) and  the Dynkin diagram
 from Fig.~1.

\begin{center}
\begin{picture}(90,20)
\put(5,5){\line(1,0){80}} \put(5,5){\circle*{1}}
\put(15,5){\circle*{1}} \put(25,5){\circle*{1}}
\put(35,5){\circle*{1}} \put(45,5){\circle*{1}}
\put(55,5){\circle*{1}} \put(65,5){\circle*{1}}
\put(75,5){\circle*{1}} \put(85,5){\circle*{1}}
\put(65,5){\line(0,1){10}} \put(65,15){\circle*{1}}
\put(5,2){\makebox(0,0)[lc]{1}} \put(15,2){\makebox(0,0)[lc]{2}}
\put(25,2){\makebox(0,0)[lc]{3}} \put(35,2){\makebox(0,0)[lc]{4}}
\put(45,2){\makebox(0,0)[lc]{5}} \put(55,2){\makebox(0,0)[lc]{6}}
\put(65,2){\makebox(0,0)[lc]{7}} \put(75,2){\makebox(0,0)[lc]{8}}
\put(85,2){\makebox(0,0)[lc]{9}}
\put(68,15){\makebox(0,0)[lc]{10}}
\end{picture} \\[5pt]
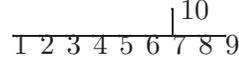
 \small Fig. 1. \it Dynkin diagram for $E_{10}$ hyperbolic KM
  algebra
 \end{center}

 Now we present a cosmological $S$-brane solution from Subsection~\ref{section4.2.1} for the conf\/iguration
 of ten branes under consideration. In what follows the relations
 $\eps_s = +1$ and $h_s = 1/2$, $s \in S$, are used.

 The metric (\ref{4.3.3}) reads:
  \begin{gather*} \nonumber
  g= \Biggl[ \bigg( \prod_{s \neq 7, 10} f_s\bigg)^{4} f_7^5 f_{10}^{11} \Biggr]^{1/14}
  \Biggl\{ - e^{2c^0 t + 2\bar  c^0}  dt \otimes dt
   + (f_2 f_4 f_6 f_9 f_{10})^{-1} e^{2c^1 t + 2\bar  c^1} dx^1 \otimes dx^1
  \\ \nonumber
 \phantom{g=}{}+ (f_4 f_7  f_{10})^{-1} e^{2c^2 t + 2\bar  c^2} dx^2 \otimes dx^2
  + (f_1 f_4 f_8  f_{10})^{-1} e^{2c^3 t + 2\bar  c^3} dx^3 \otimes dx^3
  \\ \nonumber
\phantom{g=}{} + (f_1 f_6  f_{10})^{-1} e^{2c^4 t + 2\bar  c^4} dx^4 \otimes dx^4
  + (f_5 f_{10})^{-1} e^{2c^5 t + 2\bar  c^5} dx^5 \otimes dx^5
  \\ \nonumber
 \phantom{g=}{} + (f_2 f_5 f_8 f_{10})^{-1} e^{2c^6 t + 2\bar  c^6} dx^6 \otimes dx^6
  + (f_2 f_7 f_{10})^{-1} e^{2c^7 t + 2\bar  c^7} dx^7 \otimes  dx^7
  \\ \nonumber
 \phantom{g=}{} + (f_3 f_6 f_8 f_{10})^{-1} e^{2c^8 t + 2\bar  c^8} dx^8 \otimes dx^8
  + (f_3 f_{10})^{-1} e^{2c^9 t + 2\bar  c^9} dx^9 \otimes  dx^9
  \\ \nonumber
 \phantom{g=}{} + (f_1 f_3 f_5 f_7 f_9 f_{10})^{-1} e^{2c^{10} t + 2\bar  c^{10}} dx^{10} \otimes dx^{10}
  + (f_9 f_{10})^{-1} e^{2c^{11} t + 2\bar  c^{11}} dx^{11} \otimes dx^{11}
  \\ \nonumber
 \phantom{g=}{} + \left(\prod_{s =1}^{9} f_s\right)^{-1} e^{2c^{12} t + 2\bar  c^{12}} dx^{12} \otimes dx^{12}
  +  f_7^{-1} e^{2c^{13} t + 2\bar  c^{13}}  dx^{13} \otimes dx^{13}
     \\ \nonumber
 \phantom{g=}{} +  e^{2c^{14} t + 2\bar  c^{14} } dx^{14} \otimes  dx^{14}
  +  e^{2c^{15} t + 2\bar  c^{15} } dx^{15} \otimes  dx^{15}
                     \Biggr\}.
  \end{gather*}

   For scalar f\/ields (\ref{4.3.4}) we get
  \[   \varphi^\alpha = \frac{1}{2}
  \left[ - \lambda_{5 \alpha} \left( \sum_{s \neq
  7,10} \ln f_s \right) - \lambda_{6 \alpha} \ln f_7
  + \lambda_{4 \alpha} \ln f_{10}  \right]
  + c^\alpha_{\varphi} t +\bar c^\alpha_{\varphi},  \qquad \alpha=1,\dots,5. \]
Here we used the relations $\lambda_{a}^{\alpha} = - \lambda_{a
  \alpha}$.

  The form f\/ields (see (\ref{4.3.19c}) and (\ref{4.3.19d})) are
  as follows
   \begin{gather}
   F^4 = Q_{10} dx^{12} \wedge dx^{13} \wedge dx^{14}\wedge
   dx^{15},  \nonumber  \\  \nonumber
   F^5 =
   Q_{1} f_1^{-2} f_2 dt \wedge dx^{3} \wedge dx^{4}\wedge dx^{10} \wedge dx^{12}
   + Q_{2} f_1 f_2^{-2} f_3 dt \wedge dx^{1} \wedge dx^{6}\wedge dx^{7} \wedge dx^{12}
   \\ \nonumber
     \phantom{F^4 = }{}  + Q_{3} f_2 f_3^{-2} f_4 dt \wedge dx^{8} \wedge dx^{9}\wedge dx^{10} \wedge dx^{12}
      + Q_{4} f_3 f_4^{-2} f_5 dt \wedge dx^{1} \wedge dx^{2}\wedge dx^{3} \wedge dx^{12}
   \\ \nonumber
   \phantom{F^4 = }{}+ Q_{5} f_4 f_5^{-2} f_6 dt \wedge dx^{5} \wedge dx^{6}\wedge dx^{10} \wedge dx^{12}
      + Q_{6} f_5 f_6^{-2} f_7 dt \wedge dx^{1} \wedge dx^{4}\wedge dx^{8} \wedge dx^{12}
   \\
  \phantom{F^4 = }{} + Q_{8} f_7 f_8^{-2} f_9 dt \wedge dx^{3} \wedge dx^{6}\wedge dx^{8} \wedge dx^{12}
      + Q_{9} f_8 f_9^{-2} dt \wedge dx^{1} \wedge dx^{10}\wedge dx^{11} \wedge
      dx^{12},   \nonumber   
   \\ \label{4.3.e10.f6}
   F^6 =  Q_{7} f_6 f_7^{-2} f_8 f_{10} dt \wedge dx^{2} \wedge dx^{7}\wedge dx^{10}
          \wedge dx^{12} \wedge dx^{13},
   \end{gather}
   where $Q_s \neq 0$, $s = 1, \dots, 10$.
   Here
   \[   c^0 = \sum_{j=1}^{15}  c^j,
   \qquad   \bar  c^0 = \sum_{j=1}^{15}  \bar c^j, \]
      $f_s = \exp(- q^s(t))$ and
    $q^s(t)$ obey  Toda-type equations
   \[  \ddot{q}^s = - 2 Q_s^2 \exp\left( \sum_{s' = 1}^{10} A_{s s'}
       q^{s'}\right),  \qquad s = 1, \dots, 10,\]
  where $(A_{s s'})$ is the Cartan matrix for the KM algebra $E_{10}$
  (with the Dynkin diagram  from Fig.~1)
   and the energy integration constant
   \[   E_{T} = \frac{1}{8}  \sum_{s,s' =1}^{10} A_{s s'}
   \dot{q^s} \dot{q^{s'}}
  + \frac{1}{2} \sum_{s =1}^{10}  Q_s^2  \exp\left( \sum_{s' = 1}^{10} A_{s s'} q^{s'}\right),
   \]
  obeys the constraint
  \[  2E_{T} - \sum_{\alpha = 1}^{5} (c^{\alpha}_{\varphi})^2  +  \sum_{i=1}^{15} (c^i)^2
  - \left(\sum_{i=1}^{15}  c^i \right)^2 = 0. \]

  The brane constraints (\ref{3.2.47}) are in our case
  \begin{alignat*}{3}
 &  U^1(c)= c^3 + c^4 + c^{10} + c^{12} -  \sum_{\alpha = 1}^5
  \lambda_{5 \alpha}  c^{\alpha}_{\varphi}=0, \qquad  && U^1(\bar c) = 0,&
\\
&  U^2(c)= c^1 + c^6 + c^{7} + c^{12} -  \sum_{\alpha = 1}^5
  \lambda_{5 \alpha} c^{\alpha}_{\varphi}=0,  \qquad &&  U^2(\bar c) = 0, &
\\
&   U^3(c)= c^8 + c^9 + c^{10} + c^{12} -  \sum_{\alpha = 1}^5
 \lambda_{5 \alpha} c^{\alpha}_{\varphi}=0,  \qquad  && U^3(\bar c) = 0, &
 \\
& U^4(c)= c^1 + c^2 + c^{3} + c^{12} -  \sum_{\alpha = 1}^5
 \lambda_{5 \alpha} c^{\alpha}_{\varphi}=0,  \qquad  && U^4(\bar c) = 0,&
 \\
 & U^5(c)= c^5 + c^6 + c^{10} + c^{12} -  \sum_{\alpha = 1}^5
 \lambda_{5 \alpha} c^{\alpha}_{\varphi}=0,  \qquad  && U^5(\bar c) = 0,&
 \\
& U^6(c)= c^1 + c^4 + c^{8} + c^{12} -  \sum_{\alpha = 1}^5
 \lambda_{5 \alpha} c^\alpha_{\varphi}=0,  \qquad  && U^6(\bar c) = 0,&
 \\ \nonumber
& U^7(c)= c^2 + c^7 + c^{10} + c^{12} + c^{13} -  \sum_{\alpha = 1}^5
 \lambda_{6 \alpha} c^\alpha_{\varphi}=0,  \qquad && U^7(\bar c) = 0,&
 \\
 & U^8(c)= c^3 + c^6 + c^{8} + c^{12} -  \sum_{\alpha = 1}^5
 \lambda_{5 \alpha} c^\alpha_{\varphi}=0,  \qquad  && U^8(\bar c) = 0,&
 \\
& U^9(c)= c^1 + c^{10} + c^{11} + c^{12} -  \sum_{\alpha = 1}^5
 \lambda_{5 \alpha} c^\alpha_{\varphi}=0,  \qquad  && U^{9}(\bar c) = 0,&
 \\
 & U^{10}(c)= \sum_{i = 1}^{11} c^i +  \sum_{\alpha = 1}^5
 \lambda_{4 \alpha} c^\alpha_{\varphi}=0,  \qquad  &&  U^{10}(\bar c) =
 0.  &
 \end{alignat*}

 \begin{remark}
  For a special choice of integration constants
  $c^i = 0$ and $c^{\alpha}_{\varphi} =0$, we get a solution
  governed by  $E_{10}$ Toda chain with the energy constraint
  $E_T =0$. According to the result from~\cite{IMb2a} we obtain a never
  ending asymptotical oscillating behavior of scale factors
  which is described by the
  motion of a point-like particle in a billiard $B \subset H^9$.
  This billiard has   a f\/inite volume  since $E_{10}$ is hyperbolic.
  \end{remark}

 {\bf Special 1-block solution.} Now we consider a special 1-block
 solution (see Subsection~\ref{section4.2.3}). This solution is valid when
 a special set of charges is considered (see (\ref{4.3.25n})):
  \[  Q_s^2 = Q^2 |b_s|, \]
  where $Q \neq 0$ and  \cite{GrI}
  \[   b_s = 2 \sum_{s' =1}^{10} A^{ss'} = -60, -122, -186, - 252,
   -320, -390, -462, -306, -152, -230, \]
  $s = 1,\dots, 10$.
 Recall  that $(A^{ss'})= (A_{ss'})^{-1}$.

 In this case $f_s = (\bar{f})^{b_s}$, where
 \begin{gather}    \label{4.3.e10.f}
 \bar{f}(t)=
\left\{ \begin{array}{ll}
  |Q| \sqrt{2/C} \sinh(\sqrt{C}(t- t_0)),
      &  C > 0, \vspace{1mm}\\ \nonumber
  |Q| \sqrt{2/|C|} \sin(\sqrt{|C|}( t - t_0)), &
          C < 0,  \vspace{1mm}\\ \nonumber
    |Q| \sqrt{2} ( t - t_0) , & C = 0
  \end{array}\right.
 \end{gather}
 and  $t_0$ is a constant.

 From (\ref{4.3.20e}) we get
  \[   E_{T} = -620 C, \]
 where relation $\sum_{s = 1}^{10} b_s = - 2480$
 was used.

  For the special solution under consideration the electric monomials
  in 
  (\ref{4.3.e10.f6}) have a  simpler form
    \[ {\cal F}^s= Q_s  \bar{f}^{-2}  dt \wedge\tau(I_s), \]
    where $s = 1,2,\dots,9$.

  {\bf Solution with one harmonic function.}
  Let $C= 0$ and all $c^i = \bar c^i = 0$,
   $c^\alpha_{\varphi} = \bar c^\alpha_{\varphi} = 0$. In this
  case $H = \bar{f}(t)= |Q| \sqrt{2} ( t - t_0) > 0 $ is a harmonic
  function on the 1-dimensional manifold  $((t_0, + \infty), - dt \otimes dt)$
  and  our   solution  coincides with the 1-block solution
  (\ref{3.1.11})--(\ref{3.1.17}) (if
  ${\rm sign} \nu_s = - {\rm sign}  Q_s$ for all $s$).
\end{example}

\begin{example} {\bf $\boldsymbol{S}$-brane solution governed by
   $\boldsymbol{HA_{2}^{(1)}}$ Toda chain.}
  Now we consider the $B_{11}$-model
  in  $11$-dimensional pseudo-Euclidean space of
  signature $(-, +$, $\dots, +)$
   with  4-form~$F^4$.

  Here we deal with four electric branes ($SM2$-branes)
  $s_1$, $s_2$, $s_3$, $s_4$ corresponding to the 4-form $F^4$.
  The brane sets are the following ones:
  $I_1 = \{ 1, 2, 3  \}$,
  $I_2 = \{ 4, 5, 6 \}$,
  $I_3 = \{ 7, 8, 9  \}$,
  $I_4 = \{ 1, 4, 10 \}$.

 It may be verif\/ied that these sets obey
 the intersection rules corresponding to
 the hyperbolic KM algebra $HA_{2}^{(1)}$
 with the following Cartan matrix
  \begin{gather}     \label{5.H.A.2}
   A=\barr{cccc}
    2 & -1  & -1 &  0  \\
   -1 &  2  & -1 &  0  \\
   -1 & -1  &  2 & -1  \\
    0 &  0  & -1 &  2  \\
    \earr
    \end{gather}
 (see (\ref{3.1.2.3.sl}) with $I_{s_i} = I_i$).

 Now we give a cosmological $S$-brane solution from
 Subsection~\ref{section4.2.1} for the conf\/iguration
 of four branes under consideration. In what follows the relations
 $\eps_s = +1$ and $h_s = 1/2$, $s \in S$, are used.

 The metric (\ref{4.3.3}) reads:
  \begin{gather*}  \nonumber
  g= (f_1 f_2 f_{3} f_4)^{1/3}
  \biggl\{ - e^{2c^0 t + 2\bar  c^0}  dt \otimes dt
   + (f_1 f_4)^{-1} e^{2c^1 t + 2\bar  c^1} dx^1 \otimes dx^1+ f_1^{-1} e^{2c^2 t + 2\bar  c^2} dx^2 \otimes dx^2\!
  \\ \nonumber
  \phantom{g=}{}
  + f_1^{-1} e^{2c^3 t + 2\bar  c^3} dx^3 \otimes dx^3
   + (f_2 f_4)^{-1} e^{2c^4 t + 2\bar  c^4} dx^4 \otimes dx^4
  +  f_2^{-1} e^{2c^5 t + 2\bar  c^5} dx^5 \otimes dx^5
  \\ \nonumber
 \phantom{g=}{} + f_2^{-1} e^{2c^6 t + 2\bar  c^6} dx^6 \otimes dx^6
  + f_3^{-1} e^{2c^7 t + 2\bar  c^7} dx^7 \otimes  dx^7
  + f_3^{-1} e^{2c^8 t + 2\bar  c^8} dx^8 \otimes dx^8\\
  \phantom{g=}{}
  + f_3^{-1} e^{2c^9 t + 2\bar  c^9} dx^9 \otimes  dx^9
  + f_4^{-1} e^{2c^{10} t + 2\bar  c^{10}} dx^{10} \otimes dx^{10}
    \biggr\}.  
  \end{gather*}

  The form f\/ield (see (\ref{4.3.19c})) is as   follows
   \begin{gather*}
   F^4 =
       Q_{1} f_1^{-2} f_2 f_3 dt \wedge dx^{1} \wedge dx^{2}\wedge dx^{3}
   + Q_{2} f_1 f_2^{-2} f_3 dt \wedge dx^{4} \wedge dx^{5}\wedge dx^{6}
 \\
  \phantom{F^4 =}{}  + Q_{3} f_1 f_2 f_3^{-2} dt \wedge dx^{7} \wedge dx^{8}\wedge dx^{9}
    + Q_{4} f_3 f_4^{-2} dt \wedge dx^{1} \wedge dx^{4}\wedge dx^{10},
    \end{gather*}
   where $Q_s \neq 0$, $s = 1, \dots, 4$. Here
   \[   c^0 = \sum_{j=1}^{10}  c^j,
   \qquad   \bar  c^0 = \sum_{j=1}^{10}  \bar c^j, \]
    $f_s = \exp(- q^s(t))$ and  $q^s(t)$ obey the  Toda-type equations
   \[  \ddot{q^s} = - 2 Q_s^2 \exp\left( \sum_{s' = 1 }^{4} A_{s s'}
   q^{s'}\right),   \]
  $s = 1, \dots, 4$,  where $(A_{s s'})$ is the Cartan matrix (\ref{5.H.A.2})
  for the KM algebra $HA_2^{(1)}$ and the energy integration constant
   \[
   E_{T} = \frac{1}{8}  \sum_{s,s' =1}^{4} A_{s s'}
   \dot{q^s} \dot{q^{s'}}
   + \frac{1}{2} \sum_{s =1}^{4}  Q_s^2  \exp\left( \sum_{s' = 1}^{4} A_{s s'} q^{s'} \right),
   \]
   obeys the constraint
  \[   2E_{T}  +  \sum_{i=1}^{10} (c^i)^2
  - \left(\sum_{i=1}^{10}  c^i \right)^2 = 0. \]

  The brane constraints (\ref{3.2.47}) read in this case as follows
  \begin{gather*}       \nonumber
 U^1(c)= c^1 + c^2 + c^{3} = 0, \qquad  U^1(\bar c) = 0,
 \\ \nonumber
 U^2(c)= c^4 + c^5 + c^{6} = 0,  \qquad  U^2(\bar c) = 0,
 \\ \nonumber
 U^3(c)= c^7 + c^8 + c^{9} = 0,  \qquad  U^3(\bar c) = 0,
 \\ \nonumber
 U^4(c)= c^1 + c^4 + c^{10} =0,  \qquad  U^4(\bar c) = 0.
  \end{gather*}

Since $F^4 \wedge F^4 = 0$ this solution  also obeys equations of
motion of 11-dimensional supergravity.

 {\bf Special 1-block solution.} Now we consider a special 1-block
 solution (see subsection 4.2.3). This solution is valid when
 a special set of charges is considered (see (\ref{4.3.25n})):
  \[   Q_s^2 = Q^2 |b_s|,  \]
  where $Q \neq 0$ and
   \[  b_s = 2 \sum_{s' =1}^{4} A^{ss'} = -12, -12, -14, - 6.   \]
 In this case $f_s = (\bar{f})^{b_s}$, where
  $\bar{f}$ is the same as in (\ref{4.3.e10.f}).

 For the energy integration constant we have
 \[   E_{T} = - 11 C,   \]
  (see  (\ref{4.3.20e})).
\end{example}

\begin{example}\label{example5} {\bf $\boldsymbol{S}$-brane solution governed by
 $\boldsymbol{P_{10}}$ Toda chain with $\boldsymbol{E_T = 0}$.}
  Now we consider the $B_{11}$-model
  in  $11$-dimensional pseudo-Euclidean space of
  signature $(-, +$, $\dots, +)$    with  4-form $F^4$.

  Here we deal with ten electric branes ($SM2$-branes)
  $s_1, \dots, s_{10}$ corresponding to the 4-form~$F^4$.
  The brane sets are taken from \cite{HLPS-1,HPS} as:
  $I_1 = \{ 1, 4, 7  \}$,
  $I_2 = \{ 8, 9, 10 \}$,
  $I_3 = \{ 2, 5, 7  \}$,
  $I_4 = \{ 4, 6, 10 \}$,
  $I_5 = \{ 2, 3, 9  \}$,
  $I_6 = \{ 1, 2, 8  \}$,
  $I_7 = \{ 1, 3, 10 \}$,
  $I_8 = \{ 4, 5, 8  \}$,
  $I_9 = \{ 3, 6, 7  \}$,
 $I_{10} = \{ 5, 6, 9 \}$.

 These sets obey  the intersection rules corresponding to
 the Lorentzian KM algebra $P_{10}$ (we call it Petersen algebra)
 with the following Cartan matrix
   \begin{gather}                  \label{5.P.A.2}
   A=\barr{cccccccccc}
    2 & -1  &  0 &  0 & -1 &  0 &  0 &  0  &  0 & -1 \\
   -1 &  2  & -1 &  0 &  0 &  0 &  0 &  0  & -1 &  0 \\
    0 & -1  &  2 & -1 &  0 &  0 & -1 &  0  &  0 &  0 \\
    0 &  0  & -1 &  2 & -1 & -1 &  0 &  0  &  0 &  0 \\
   -1 &  0  &  0 & -1 &  2 &  0 &  0 & -1  &  0 &  0 \\
    0 &  0  &  0 & -1 &  0 &  2 &  0 &  0  & -1 & -1 \\
    0 &  0  & -1 &  0 &  0 &  0 &  2 & -1  &  0 & -1 \\
    0 &  0  &  0 &  0 & -1 &  0 & -1 &  2  & -1 &  0 \\
    0 & -1  &  0 &  0 &  0 & -1 &  0 & -1  &  2 &  0 \\
   -1 &  0  &  0 &  0 &  0 & -1 & -1 &  0  &  0 &  2
    \earr.
    \end{gather}

The Dynkin diagram for this Cartan matrix could be represented by
the Petersen graph  (``a~star inside a pentagon''). $P_{10}$ is
the Lorentzian KM algebra. It is a subalgebra of $E_{10}$
\cite{HLPS-1,HPS}.

 Let us present an $S$-brane solution  for the conf\/iguration
 of $10$ electric branes under consideration.
 The metric (\ref{4.3.3}) reads:
  \begin{gather*}
  g= \left(\prod_{s =1}^{10} f_s \right)^{1/3}
  \biggl\{ -  dt \otimes dt
   + (f_1 f_6 f_7)^{-1}  dx^1 \otimes dx^1
  + (f_3 f_5 f_6)^{-1}  dx^2 \otimes dx^2\\
  \phantom{g=}{}
  + (f_5 f_7 f_9)^{-1}  dx^3 \otimes dx^3
   +  (f_1 f_4 f_8)^{-1} dx^4 \otimes dx^4
  +  (f_3 f_8 f_{10})^{-1}      dx^5 \otimes dx^5
  \\ \nonumber
\phantom{g=}{}  +  (f_4 f_9 f_{10})^{-1}  dx^6 \otimes dx^6
  +  (f_1 f_3 f_9)^{-1}  dx^7 \otimes  dx^7
  +  (f_2 f_6 f_8)^{-1}   dx^8 \otimes dx^8\\
\phantom{g=}{}   + (f_2 f_5 f_{10})^{-1}   dx^9 \otimes  dx^9
  + (f_2 f_4 f_7)^{-1}   dx^{10} \otimes dx^{10}
    \biggr\}.  \label{5.3.p.g}
  \end{gather*}

  The form f\/ield (see (\ref{4.3.19c})) is
  the following
   \begin{gather*}
   F^4 =
    Q_{1} f_1^{-2} f_2 f_5 f_{10} dt \wedge dx^{1} \wedge dx^{4}\wedge dx^{7}
   + Q_{2} f_1 f_2^{-2} f_3 f_9   dt \wedge dx^{8} \wedge dx^{9}\wedge dx^{10}
   \nonumber
   \\
   \phantom{F^4 =}{}
    + Q_{3} f_2 f_3^{-2} f_4 f_7 dt \wedge dx^{2} \wedge dx^{5}\wedge dx^{7}
    + Q_{4} f_3 f_4^{-2} f_5 f_6 dt \wedge dx^{4} \wedge dx^{6}\wedge  dx^{10}
    \\
   \phantom{F^4 =}{}+
    Q_{5} f_1 f_4 f_5^{-2} f_8  dt \wedge dx^{2} \wedge dx^{3}\wedge dx^{9}
  + Q_{6} f_4  f_6^{-2} f_9 f_{10} dt \wedge dx^{1} \wedge dx^{2}\wedge dx^{8}
   \\
   \phantom{F^4 =}{}
    + Q_{7} f_3  f_7^{-2} f_8 f_{10} dt \wedge dx^{1} \wedge dx^{3}\wedge dx^{10}
    + Q_{8} f_5 f_7 f_8^{-2} f_9     dt \wedge dx^{4} \wedge dx^{5}\wedge dx^{8}
    \\
    \phantom{F^4 =}{}+
    Q_{9} f_2 f_6 f_8 f_9^{-2}     dt \wedge dx^{3} \wedge dx^{6}\wedge dx^{7}
   + Q_{10} f_1 f_6 f_7 f_{10}^{-2}  dt \wedge dx^{5} \wedge dx^{6}\wedge dx^{9},
     \end{gather*}
   where $Q_s \neq 0$, $s = 1, \dots, 10$. Here
  $f_s = \exp(- q^s(t))$ and   $q^s(t)$ obey  the Toda-type equations
   \[    \ddot{q}^s = - 2 Q_s^2 \exp\left( \sum_{s' = 1 }^{10} A_{s s'}
   q^{s'} \right), \]
  where $(A_{s s'})$ is the Cartan matrix (\ref{5.P.A.2})
  for the KM algebra $P_{10}$ and the  energy constraint
   \[   E_{T} = \frac{1}{8}  \sum_{s,s' =1}^{10} A_{s s'}
  \dot{q}^s \dot{q}^{s'}
  + \frac{1}{2} \sum_{s =1}^{10}
  Q_s^2  \exp\left( \sum_{s' = 1}^{10} A_{s s'} q^{s'}\right) = 0 \]
  is obeyed. Here we used the fact that the  two sets of
  linear equations    $U^s(c)= 0$,  $U^s(\bar c) = 0$, $s =1,\dots,
  10$,  have trivial solutions: $c= 0$, $\bar c= 0$, due to
  the linear independence of vectors~$U^s$.

 Since $F^4 \wedge F^4 = 0$, this solution  also obeys the equations
 of motion of 11-dimensional supergravity.
 \end{example}

  \begin{remark}
  As pointed out in \cite{HLPS-1}
   we do not obtain  a never
  ending asymptotic oscillating behavior of the scale factors
  in this case since the Lorentzian KM algebra $P_{10}$
  is not hyperbolic and  the corresponding billiard $B \subset H^9$  has
   an inf\/inite volume.
   \end{remark}

{\bf Special 1-block solution.} Now we consider a special 1-block
 solution. The  calculations give us the following relations
  \[  b_s = 2 \sum_{s' =1}^{10} A^{ss'} = -2,\qquad s =1,\dots,  10, \]
 and hence the special solution is valid (see (\ref{4.3.25n})), when
 all charges are equal
 \[  Q_s^2 = Q^2, \]
  where $Q \neq 0$.
 In this case all $f_s = \bar{f}^{-2}$, where
 \[  \bar{f}(t)=  |Q| (t - t_0), \]
  and $t_0$ is constant.  The metric (\ref{5.3.p.g}) may be  rewritten using
 the synchronous time variable $t_s$:
  \[  g= - dt_s \otimes dt_s + A t_s^{2/7} \sum_{i=1}^{10} dx^i \otimes dx^i,
  \]
 where  $A > 0$ and $t_s > 0$. This metric coincides with the
 power-law, inf\/lationary solution in the model with a one-component
 perfect f\/luid when the following equation of state is adopted:
 $p = \frac{2}{5} \rho$, where $p$ is pressure and $\rho$ is
 the density of f\/luid \cite{IM-pf,AIKM-pf}.

 \section{Black brane solutions}\label{section5}

 In this section we consider the spherically symmetric case
of the metric (\ref{4.3.19}), i.e.\ we put $w = 1$, $M_1 =
  S^{d_1}$, $g^1 = d \Omega^2_{d_1}$, where $d \Omega^2_{d_1}$ is
the canonical metric on a unit sphere $S^{d_1}$, $d_1 \geq 2$. In
this case $\xi^1 = d_1 -1$. We put  $M_2 = \R$, $g^2 = - dt
\otimes dt$, i.e.\  $M_2$ is a time manifold.

Let $C_1 \geq 0$. We consider   solutions def\/ined on some interval
$[u_0, +\infty)$ with a  horizon at $u = + \infty$.

When the matrix $(h_{\alpha\beta})$ is positive
def\/inite  and
 \[   2 \in I_s, \qquad \forall \, s \in S, \]
i.e.\ all branes have a common time direction $t$, the horizon
condition  singles out the unique solution with $C_1
  > 0$ and linear asymptotics at inf\/inity
 \[  q^s = -  \beta^s u + \bar \beta^s  + o(1), \]
 $u \to +\infty$, where
 $\beta^s, \bar \beta^s$ are constants, $s \in S$
 \cite{IMp2,IMp3}.

In this case
  \begin{gather}
 c^A/\bar{\mu}  = - \delta^{A}_{2} + h_1 U^{1 A}  +
 \sum_{s\in S}  h_s b_s U^{s A},  \qquad
    \label{5.2.20}
 \beta^s/\bar{\mu} = 2 \sum_{s' \in S} A^{s s'} \equiv b_s,
 \end{gather}
where $s \in S$, $A = (i, \alpha)$, $\bar{\mu} = \sqrt{C_1}$, the
matrix $(A^{s s'})$ is inverse of the generalized Cartan matrix
$(A_{s s'})$ and  $h_1 = (U^1, U^1)^{-1} = d_1/(1 - d_1)$.

Let us introduce a new radial variable $R = R(u)$ through the
relations
 \[ \exp( - 2\bar{\mu} u) = 1 -
 \frac{2\mu}{R^{\bar{d}}}, \qquad \mu = \bar{\mu}/ \bar{d} >0,
 \]
where $u > 0$, $R^{\bar d} > 2\mu$, $\bar d = d_1 -1$. We put
 $\bar{c}^A = 0$ and $q^s(0) = 0$,
 $A = (i, \alpha)$, $s \in S$. These relations guarantee the
asymptotic f\/latness (for $R \to +\infty$) of the
 $(2+d_1)$-dimensional section of the metric.

Let us denote  $H_s = f_s e^{- \beta^s u }$, $s \in S$. Then,
solutions (\ref{4.3.19})--(\ref{4.3.19b}) may be written as follows
 \cite{IMp1,IMp2,IMp3}
 \begin{gather}
 g= \left(\prod_{s \in S}
  H_s^{2 h_s d(I_s)/(D-2)} \right) \left\{ \left(1 -
 \frac{2\mu}{R^{\bar{d}}}\right)^{-1} dR \otimes dR + R^2  d
 \Omega^2_{d_1}  \right.\nonumber  \\  \label{5.2.30}
\left.\phantom{g=}{}  -  \left(\prod_{s \in S} H_s^{-2 h_s}
 \right) \left(1 - \frac{2\mu}{R^{\bar{d}}}\right) dt \otimes dt +
 \sum_{i = 3}^{n} \left(\prod_{s \in S}
  H_s^{-2 h_s \delta_{iI_s}} \right) \hat{g}^i  \right\},
 \\  \label{5.2.31}
 \exp(\varphi^\alpha)=
 \prod_{s\in S} H_s^{h_s \chi_s \lambda_{a_s}^\alpha},
 \end{gather}
 where
 $F^a= \sum_{s \in S} \delta^a_{a_s} {\cal F}^{s}$, and
 \begin{gather}          \label{5.2.32}
 {\cal F}^s= - \frac{Q_s}{R^{d_1}} \left( \prod_{s'
 \in S}  H_{s'}^{- A_{s s'}} \right) dR \wedge\tau(I_s), \qquad s\in  S_e,
\\ \label{5.2.33}
 {\cal F}^s= Q_s \tau(\bar I_s),\qquad s \in S_m.
 \end{gather}
 Here $Q_s \neq 0$, $h_s =K_s^{-1}$, $s \in S$, and the
generalized Cartan matrix $(A_{s s'})$ is non-degenerate.

Functions $H_s > 0$ obey the equations
 \begin{gather}        \label{5.3.1}
 \frac{d}{dz} \left( \frac{(1 - 2\mu z)}{H_s}
 \frac{d}{dz} H_s \right) = \bar B_s
 \prod_{s' \in S}  H_{s'}^{- A_{s s'}},
 \\       \label{5.3.2a}
  H_{s}((2\mu)^{-1} -0) = H_{s0} \in (0, + \infty), \\
  \label{5.3.2b} H_{s}(+ 0) = 1,\qquad s \in S,
  \end{gather}
where $H_s(z) > 0$, $\mu > 0$, $z = R^{-\bar d} \in (0,
 (2\mu)^{-1})$ and $\bar B_s = \eps_s K_s Q_s^2/ \bar d^2 \neq 0$.

There exist solutions to equations (\ref{5.3.1})--(\ref{5.3.2a}) of
polynomial type. The simplest example occurs in orthogonal case
 \cite{BIM,IMJ} (for $d_i = 1$  see also \cite{AIV,Oh}):
 $(U^s,U^{s'})= 0$, for  $s \neq s'$, $s, s' \in S$. In this case
 $(A_{s s'}) = {\rm diag}(2,\ldots,2)$ is a Cartan matrix for
 the semisimple Lie algebra ${\bf A_1} \oplus  \cdots  \oplus  {\bf
 A_1}$ and
 \begin{gather}
 H_{s}(z) = 1 + P_s z    \label{5.3.5}
 \end{gather}
 with $P_s \neq 0$, satisfying
 \begin{gather*}
 P_s(P_s + 2\mu) = -\bar B_s,\qquad s \in S.
 \end{gather*}
(For earlier supergravity solutions  see \cite{CT,OS} and
 references therein).

In \cite{Br1,IMJ2,CIM} this solution was generalized to a
block-orthogonal case  (\ref{3.1.5}), (\ref{3.1.6}). In this case
(\ref{5.3.5}) is modif\/ied as follows
  \begin{gather}     \label{5.3.8}
  H_{s}(z) = (1 +   P_s z)^{b_s},
  \end{gather}
where $b_s$ are def\/ined in  (\ref{5.2.20}) and parameters $P_s$
coincide inside blocks, i.e.\ $P_s = P_{s'}$ for $s, s'
 \in S_i$, $i =1,\dots,k$. The parameters $P_s \neq 0 $ satisfy the
relations \cite{IMJ2,CIM,IM-top}
 \begin{gather}       \label{5.3.5b}
 P_s(P_s + 2\mu) = -\bar B_s/b_s, \qquad s \in S,
 \end{gather}
 and the parameters $\bar B_s/b_s$  coincide inside
 blocks, i.e.\ $\bar B_s/b_s = \bar B_{s'}/b_{s'}$ for $s, s'
 \in S_i$, $i =1,\dots,k$.

{\bf Finite-dimensional Lie algebras.}
 Let $(A_{s s'})$ be  a
Cartan matrix  for a  f\/inite-dimensional semisimple Lie  algebra
$\cal G$. In this case all powers $b_s$ def\/ined in (\ref{5.2.20})  are  natural
numbers  which coincide with the components of twice the  dual
Weyl vector in the basis of simple co-roots \cite{FS} and  hence,
all functions $H_s$ are polynomials,
 $s \in S$.

\begin{conjecture}\label{conjecture1} Let $(A_{s s'})$ be  a Cartan matrix for a
 semisimple finite-dimensional Lie algeb\-ra~$\cal G$. Then  the
 solutions to equations \eqref{5.3.1}--\eqref{5.3.2b} $($if exist$)$ have
 a polynomial structure:
  \begin{gather*}       
  H_{s}(z) = 1 + \sum_{k = 1}^{n_s} P_s^{(k)} z^k,
  \end{gather*}
  where $P_s^{(k)}$ are constants, $k =
  1,\ldots, n_s$; $n_s = b_s =
  2 \sum_{s' \in S} A^{s s'} \in {\mathbb N}$
  and $P_s^{(n_s)} \neq 0$,  $s \in S$.
\end{conjecture}

In the extremal case ($\mu = + 0$) an analogue of this conjecture
was suggested previously in~\cite{LMMP}. Conjecture~\ref{conjecture1} was verif\/ied
for the
 ${\bf A_m}$ and ${\bf C_{m+1}}$ Lie algebras in
 \cite{IMp2,IMp3}. Explicit expressions for polynomials
 corresponding to Lie algebras $C_2$ and $A_3$ were obtained in~\cite{GrIK} and~\cite{GrIM} respectively.

{\bf Hyperbolic KM algebras.} Let $(A_{s s'})$ be  a Cartan matrix
for an inf\/inite-dimensional hyperbolic KM  algebra $\cal G$. In
this case all powers in (\ref{5.2.20})  are  negative numbers and
hence, we have no chance to get a polynomial structure for $H_s$.
Here we are led to an open problem of seeking  solutions to the
set of ``master'' equations (\ref{5.3.1}) with boundary conditions~(\ref{5.3.2a}) and~(\ref{5.3.2b}). These solutions def\/ine special
solutions to Toda-chain equations corresponding to the hyperbolic
KM algebra~$\cal G$.

\begin{example}\label{example6}
{\bf  Black hole solutions for   $\boldsymbol{A_1 \oplus A_1}$, $\boldsymbol{A_2}$
and  $\boldsymbol{H_2(q,q)}$ KM algebras.}
  Let us consider the 4-dimensional model governed by the
  action
   \[  S =  \int_{M} d^{4}z \sqrt{|g|} \left\{ {R}[g] -  \eps
     g^{MN} \partial_{M} \varphi\partial_{N} \varphi
    -  \frac{1}{2} e^{2\lambda \varphi} (F^1)^2
    -  \frac{1}{2} e^{- 2\lambda \varphi} (F^2)^2\right\}.  \]
 Here $F^1$ and $F^2$ are 2-forms, $\varphi$ is scalar f\/ield
  and $\eps = \pm 1$.

  We consider a black brane solution
  def\/ined on $\R_{*} \times S^{2} \times \R $
  with two electric branes
  $s_1$ and $s_2$ corresponding to forms $F^1$ and $F^2$,
  respectively, with the sets $I_1 = I_2 = \{ 2 \}$.
  Here $\R_{*}$ is subset of $\R$,
  $M_1 = S^{2}$,  $g^1 = d \Omega^2_{2}$,
  is the canonical metric on $S^{2}$, $M_2 = \R$,
  $g^2 = - dt \otimes dt$ and $\eps_1 = \eps_2 = -1$.

  The scalar products of $U$-vectors are (we identify $U^i = U^{s_i}$):
  \[   (U^1,U^1) = (U^2, U^2) = \frac{1}{2 } + \eps \lambda^2  \neq 0,
   \qquad (U^1,U^2) = \frac{1}{2 } - \eps \lambda^2.  \]

    The matrix $A$ from (\ref{3.1.2.2})
   is a generalized non-degenerate Cartan matrix  if and only if
   \[    \frac{2(U^1,U^2)}{(U^2, U^2)} = -q, \]
   or, equivalently,
   \[   \eps \lambda^2 = \frac{2+q}{2(2-q)}, \]
      where $q = 0,1,3,4, \dots$. This takes place
   when
   \begin{gather*}
    \eps = + 1, \qquad q = 0, 1, \qquad
       \eps = - 1, \qquad q =  3, 4, 5, \dots
    \end{gather*}
    and
     \[ \lambda^2 = \frac{2+q}{2 |2-q|}. \]

    The f\/irst branch ($\eps = + 1$) corresponds to f\/inite
    dimensional Lie algebras $A_1 \oplus A_1$ ($q =0$),
    $A_2$ ($q = 1$)
     and   the second one ($\eps = - 1$) corresponds to
    hyperbolic KM algebras $H_2(q,q)$, $q = 3,4, \dots$.
    In the hyperbolic case the scalar f\/ield $\varphi$
    is a phantom (ghost).

    The  black brane solution reads (see
    (\ref{5.2.30})--(\ref{5.2.32}))
    \begin{gather}
    g= (H_1 H_2)^{h} \left\{ \left(1 -
   \frac{2\mu}{R} \right)^{-1} dR \otimes dR + R^2  d \Omega^2_{2}
         - (H_1 H_2)^{- 2h}
     \left(1 - \frac{2\mu}{R}\right) dt \otimes dt \right\},
      \label{5.4.7}     \\  \label{5.4.8}
   \exp(\varphi)= (H_1 / H_2)^{\eps \lambda h},
    \\  \label{5.4.9}
    F^s= \frac{Q_s}{R^2} H_s^{-2} (H_{\bar s})^{q} dt \wedge dR,\qquad s = 1, 2.
    \end{gather}
 Here $h = (2- q)/2$ and $\bar s = 2, 1 $ for $s = 1, 2$
    respectively.

    The moduli functions $H_s > 0$ obey
   the equations (see (\ref{5.3.1}))
  \begin{gather}         \label{5.4.10}
  \frac{d}{dz} \left( \frac{(1 - 2\mu z)}{H_s}
  \frac{d}{dz} H_s \right) = \frac{2 Q_s^2}{q-2} H_s^{-2} (H_{\bar s})^{q},
  \end{gather}
  with the boundary conditions  $H_{s}((2\mu)^{-1} -0) = H_{s0} \in (0, + \infty)$,
  $H_{s}(+ 0) = 1$, $s = 1, 2$, imposed. Here  $\mu > 0$, $z = 1/R \in (0,
   (2\mu)^{-1})$. For $q = 0, 1$ the solutions to equations (\ref{5.4.10}) with
   the boundary conditions imposed were given in
   \cite{IMp1,IMp2,IMp3}. They are polynomials of degrees~$1$
    and~$2$ for $q=0$ and $q =1$, respectively. For $q = 3, 4,
    \dots$ the exact solutions to equations~(\ref{5.4.10}) are not known
    yet.
   \end{example}

   {\bf Special solution with $\boldsymbol{Q_1^2 = Q_2^2}$.}
   Now we consider the special one-block
   solution with the functions $H_s$
   obeying (\ref{5.3.8}) and  (\ref{5.3.5b}). Since $b_s = 2/(2 - q)$
   and $\bar B_s = 2 Q_s^2/(q - 2)$ it takes place when  $Q_1^2 = Q_2^2 =
   Q^2 > 0$. The moduli functions read
   \[   H_s  = H^{2/(2 - q)}, \qquad  H =  1 + P z, \]
    where $z = 1/R$ and $q \neq 2$.  These functions obey
    $H_s(z) > 0$ for $z \in [0, (2 \mu)^{-1}]$
    if $P > - 2 \mu$ ($\mu > 0$). Due to this inequality and the
    relation  $P(P + 2 \mu) = Q^2$ (following from (\ref{5.3.5b}) ) we get
     \[     P = -  \mu + \sqrt{\mu^2 + Q^2} > 0.  \]

     In this special case the solution (\ref{5.4.7})--(\ref{5.4.9})
     has the following form:
       \begin{gather}      g= H^{2} \left\{ \left(1 -
   \frac{2\mu}{R} \right)^{-1} dR \otimes dR + R^2  d \Omega^2_{2}
     - H^{- 4}  \left(1 - \frac{2\mu}{R}\right) dt \otimes dt \right\},\nonumber
   \\
   \varphi = 0,\label{5.4.7s}
    \\  \nonumber
    F^s= \frac{Q_s}{H^2 R^2} dt \wedge dR,\qquad s = 1, 2.
    \end{gather}
Remarkably, this special solution does not depend
    upon $q$. The metric  (\ref{5.4.7s}) coincides with the
    metric of the Reissner--Nordstr\"om solution (when the Maxwell 2-form
    is $F = \sqrt{2} Q (HR)^{-2} dt \wedge dR$).

    In the extremal case $\mu \to + 0$ we are lead to the special
    case of a  Majumdar--Papapetrou type solution
   \begin{gather*}
    g=H^2 \hat{g}^0-H^{-2}dt\otimes dt,
    \\
    \varphi = 0, \\
        \nonumber
    F^s = \nu_s dH^{-1} \wedge dt,
    \end{gather*}
   where  $g^0 = \sum_{i=1}^{3} dx^i \otimes dx^i$,
    $H$ is a harmonic function on $M_0 = \R^3$
    and $\nu_s^2 = 1$, $s = 1,2$. Here $\nu_s = - Q_s/Q$.

 \section{Conclusions}\label{section6}

Here we reviewed several  families of exact solutions in
multidimensional gravity with a set of scalar f\/ields and f\/ields of
forms  related to non-singular (e.g. hyperbolic) KM algebras.

The solutions describe composite  electromagnetic branes def\/ined
on warped products of  Ricci-f\/lat, or sometimes Einstein, spaces
of arbitrary dimensions and signatures. The metrics are
block-diagonal and all scale factors, scalar f\/ields and f\/ields of
forms depend on points of some  manifold $M_0$. The solutions
include those depending upon harmonic functions,
 S-branes and spherically-symmetric solutions (e.g. black-branes).
 Our approach is based on the  sigma-model representation
 obtained in \cite{IMC}  under the rather general assumption on
intersections of composite branes (when stress-energy tensor has a
diagonal structure).

 We were dealing with rather general intersection rules
 \cite{IMJ} governed by  invertible generalized Cartan matrix
 corresponding to the certain generalized  KM Lie
 algebra ${\cal G}$. For ${\cal G} = {\bf A_1} \oplus   \dots
 \oplus {\bf A_1}$ ($r$ terms) we get the well-known
 standard (e.g. supersymmetry preserving)
 intersection rules \cite{AR,AEH,AIR,IMC}.

 We have also considered a class of special
 ``block-orthogonal''  solutions
 corresponding to semisimple KM algebras and
 governed by several harmonic functions.
 Certain examples  of 1-block solutions (e.g. corresponding to KM algebras
 $H_2(q,q)$, $AE_3$) were  considered.

 In  the one-block case  a  generalization of the solutions
 to those governed by several functions
 of one harmonic function $H$ and obeying Toda-type equations was
 presented.

 For f\/inite-dimensional (semi-simple) Lie algebras
 we are led to integrable Lagrange systems while the
 Toda chains corresponding to inf\/inite-dimensional (non-singular)
 KM algebras are not well studied yet.
 Some examples of $S$-brane solutions corresponding to
 Lorentzian KM algebras  $HA_{2}^{(1)} = A_2^{++}$, $E_{10}$ and $P_{10}$
 were presented.

  We have also considered general classes of cosmological-type solutions
  (e.g.\  $S$-brane and spherically symmetric solutions)  governed
  by Toda-type  equations, containing black brane con\-f\/i\-gu\-ra\-tions
  as a special case. The ``master'' equations for moduli functions
  have polynomial solutions in the f\/inite-dimensional case (according
  to our conjecture \cite{IMp1,IMp2,IMp3}),
  while in the inf\/inite-dimensional case  we have only a special family of
  the so-called block-orthogonal solutions
  corresponding to semi-simple  non-singular  KM algebras.
  Examples of 4-dimensional dilatonic black hole solutions corresponding to
  KM algebras  $A_1 \oplus A_1$, $A_2$ and  $H_2(q,q)$ ($q > 2$)
  were given.

  We note that the problem of integrability
  of Toda chain equations corresponding to (non-singular)
  KM algebras  arises  also in the context of
  f\/luxbrane solutions  \cite{Iflux}
  that have also a~polynomial structure of moduli functions
  for f\/inite-dimensional Lie algebras (see also \cite{GolI}).
  (For similar $S$-brane solutions governed by polynomial functions
  and its applications in connection with cosmological
  problems see \cite{GonIM,AlIM,IKonM}.)

  Here we have  considered only the case of non-degenerate matrix $A$.
  It is an open problem to f\/ind  general classes of solutions with
  branes  for the degenerate case when ${\rm det} A = 0$ (e.g. corresponding to
   af\/f\/ine KM  algebras).  Some  special solutions of such type with maximal set of
  composite electric $S$-branes
  (e.g.\ when $A$ is not obviously a generalized Cartan matrix)
  were found in \cite{IS,IMS-pr}  and generalized in \cite{IMS,DIM}
  for arbitrary  (anti-)self-dual parallel charge density form of dimension
  $2m$ def\/ined on  Ricci-f\/lat Riemannian sub-manifold of dimension $4m$.
  In these examples the restrictions on
  brane intersections  (\ref{2.2.2a}) and (\ref{2.2.3a}) were
  replaced by more general condition on the stress-energy tensor:
   $T^M_N = 0$, $M \neq N$.

  It should be noted here that  the solutions related to
  Lorentzian (e.g. hyperbolic) KM algebras considered in Examples \ref{example3}--\ref{example6} are the new
  ones. (In Example~\ref{example5} the brane conf\/iguration from \cite{HLPS-1,HPS})
  was used.)

 \subsection*{Acknowledgements}

 This work was supported in part by the Russian Foundation for
 Basic Research grant  Nr.~07--02--13624--of\/i$_{\rm ts}$. We are
 grateful to  D.~Singleton for reading the manuscript and valuable comments.
 We are also indebted to anonymous referees whose comments have led to
 the  improvement of the paper.

\pdfbookmark[1]{References}{ref}
\LastPageEnding

 \end{document}